\documentclass{article}

\usepackage{PRIMEarxiv}

\usepackage[utf8]{inputenc}
\usepackage[T1]{fontenc}
\usepackage[hypertexnames=false]{hyperref}
\usepackage{url}
\usepackage{booktabs}
\usepackage{amsfonts}
\usepackage{amsmath}
\usepackage{nicefrac}
\usepackage{microtype}
\usepackage{fancyhdr}
\usepackage{graphicx}
\usepackage{tikz}
\usetikzlibrary{arrows.meta,positioning,shapes.geometric,fit,calc,decorations.pathreplacing}
\usepackage{enumitem}
\usepackage{algorithm}
\usepackage{algpseudocode}
\usepackage{pifont}
\usepackage{amssymb}
\usepackage{xspace}
\graphicspath{{media/}}

\newcommand{\cmark}{\ding{51}}
\newcommand{\xmark}{\ding{55}}
\newif\ifanonymous
\anonymousfalse

\ifanonymous
  \newcommand{\toolname}{{\sc BoundaryFlow}\xspace}
  \newcommand{\anonrepo}{\url{https://anonymous.4open.science/r/XXXX}}
  \newcommand{\authorblock}{%
    \author{Anonymous Submission \\ Paper ID: TBD}%
  }
\else
  \newcommand{\toolname}{{\sc kcov-dataflow}\xspace}
  \newcommand{\anonrepo}{\url{https://github.com/yskzalloc/kcov-dataflow}}
  \newcommand{\authorblock}{%
    \author{Yunseong Kim \\
      Ericsson Software Technology \\
      \texttt{yunseong.kim@est.tech}}%
  }
\fi

\newcommand{\sancov}{\texttt{SanitizerCoverage}\xspace}
\newcommand{\tracepc}{\texttt{trace-pc}\xspace}

\newcommand{\kcov}{\texttt{KCOV}\xspace}

\usepackage{listings}
\renewcommand{\shorttitle}{Per-Task Data-Flow Extraction at Kernel Function Boundaries}
\renewcommand{\headeright}{A Preprint}

\title{Beyond Edge Coverage: Per-Task Data-Flow Extraction at Kernel Function Boundaries via LLVM
}

\authorblock
\date{}

\begin{document}
\maketitle

\begin{abstract}
Coverage-guided kernel fuzzers such as syzkaller rely on edge coverage
(\texttt{trace-pc}) as their sole feedback signal. This context-blind approach
cannot distinguish execution paths that differ only in argument values---for
example, two invocations of \texttt{copy\_from\_user()} with different size
parameters hit identical basic blocks yet have vastly different security
implications. I present \toolname{}, an LLVM-based instrumentation framework
that extends Linux KCOV with per-task data-flow extraction at function
boundaries. Following a strict IR-only design rule---debug information may
enrich offline symbolization but must never determine what instrumentation is
emitted---an LLVM \sancov{} pass emits lightweight \texttt{notail},
\texttt{nomerge} three-operand callbacks on function entry and return.
A scalar, pointer address, or register-resident element is passed directly in a
register alongside a packed $(\mathit{arg\_idx}, \mathit{size})$ tag and a null
\texttt{offsets} table. When the IR itself guarantees a readable memory region---either
an ABI-in-memory object (\texttt{byval}, \texttt{byref}, \texttt{inalloca},
\texttt{preallocated}, \texttt{sret}) or a \texttt{noundef} pointer with an
explicit \texttt{dereferenceable(N)} or \texttt{dereferenceable\_or\_null(N)}
contract---the callback carries that address together with a constant
\texttt{offsets} table, up to \texttt{-sanitizer-coverage-trace-pointee-bytes}
(default~64). On an x86-64 Linux kernel build this captures pointee snapshots on
42\% of Rust object pointer arguments (318/756), while plain C pointers carry
no such IR attribute (0/23{,}810); a complementary compiler mode and
\texttt{KCOV\_DF\_TYPE\_ACCESS} (\texttt{0xA}) kernel record path close that gap
by reporting constant-offset loads and stores through pointer arguments using
zero extra memory reads. The callbacks carry no program counter---the kernel
derives it from the return address minus one---and an append-only per-task
buffer (plus bounded remote kworker collection with atomic epoch reset and
memory-pressure reclamation) delivers variable-length entry, return, access, and
comparison records without interfering with mainline KCOV. Two Rust
instrumentation paths are provided---a post-compilation IR pipeline and native
rustc instrumentation---decoding idiomatic Rust objects that \texttt{ftrace}
(BTF-bound, six-argument limit) and \texttt{drgn} (elided at \texttt{-O2})
cannot observe.
\end{abstract}


\section{Introduction}\label{sec:intro}

The Linux kernel exceeds 44 million lines of tracked source (38.4M of it C,
headers, and assembly), integrates Rust ($\sim$199k lines, 160k of them in
\texttt{rust/}) as a second
systems language, and implements complex locking hierarchies
(\texttt{rwsem}, \texttt{mutex}, RCU) that create deep state spaces unreachable
by coverage-guided fuzzing alone. syzkaller~\cite{vyukov2015syzkaller} has
discovered over 5{,}000 kernel bugs in the linux kernel, yet plateaus on stateful
subsystems such as \texttt{ksmbd}, \texttt{io\_uring}, and \texttt{binder}
where security-critical behavior depends on runtime argument values rather than
control-flow topology.

The fundamental limitation is a \emph{semantic gap}: KCOV's \tracepc reports
\emph{which} basic blocks executed but not \emph{with what values}
(Figure~\ref{fig:blind-spot}). Two
executions of \texttt{vfs\_open()} with different access-mode flags traverse
identical edges yet have entirely different security implications. Fuzzers
cannot distinguish these without data-flow context, leading to coverage
saturation on value-dependent state transitions.

Prior approaches each fall short for kernel-scale deployment. Dynamic tracing
via \texttt{ftrace}/\texttt{kprobes} imposes per-function overhead unsuitable
for continuous fuzzing. IJON~\cite{aschermann2020ijon} requires manual
annotation---infeasible for 30M~LOC. Context-sensitive coverage
(Angora~\cite{chen2018angora}, CollAFL~\cite{gan2018collafl}) clones call sites
but explodes on polymorphic dispatch (\texttt{file\_operations} alone has 20+
function pointers). DataFlowSanitizer provides byte-level taint tracking but
imposes 10--100$\times$ overhead, rendering it unusable for kernel fuzzing.

I observe that \emph{function boundaries} are natural chokepoints where
data-flow context is both cheap to capture and highly informative. Arguments
encode the \emph{intent} of a call; return values encode the \emph{outcome}.
Capturing these at compile time with a lock-free kernel transport gives both
fuzzers and analysts a new dimension of observability without source annotation.

\toolname{} is an automated LLVM IR instrumentation framework that reports every
IR argument and return value via a three-operand callback
$\langle \mathit{tag}, \mathit{val}, \mathit{offsets} \rangle$ (with
$\mathit{tag} = (\mathit{arg\_idx} \ll 32) \mid \mathit{size}$ for arguments and
$\mathit{size}$ for returns), which the kernel turns into a variable-length
record $\langle \mathit{PC}, \mathit{arg\_idx}, \mathit{size},
\mathit{address}, \mathit{values}[\,] \rangle$.
It operates through three layers:
\begin{enumerate}[nosep]
  \item A strictly IR-only compiler pass---whose emitted callbacks, operands,
        and machine code are invariant to \texttt{-g}---that reports register
        values directly and uses constant IR-derived \texttt{offsets} tables for
        ABI-in-memory objects (\texttt{byval}/\texttt{sret}) and
        IR-\texttt{dereferenceable} pointee snapshots, complemented by optional
        \texttt{KCOV\_DF\_TYPE\_ACCESS} tracking of constant-offset loads and
        stores through plain pointer arguments.
  \item A kernel-resident lock-free per-task record buffer (and bounded remote
        kworker transport with explicit epoch resets and per-CPU drop
        telemetry) delivered through \texttt{/sys/kernel/debug/kcov\_dataflow}.
  \item A user-space consumer interface for both fuzzer executors and human
        analysts, with optional offline DWARF symbolization
        (\texttt{sancov -symbolize-dataflow} / \texttt{addr2line}).
\end{enumerate}

My contributions are:
\begin{itemize}[nosep]
  \item An IR-only LLVM \sancov extension\footnote{\url{https://github.com/llvm/llvm-project/pull/201410}} (paired with Clang driver support, weak \texttt{compiler-rt} runtime stubs, and formal documentation) whose emitted callbacks depend solely on LLVM IR types and attributes---preserving identical instrumentation with or without \texttt{-g}---and automatically capture ABI-in-memory structs, IR-verified \texttt{dereferenceable} pointee snapshots, and (under an optional follow-on mode) constant-offset pointee loads/stores without extra reads.
  \item A lock-free per-task kernel transport (\texttt{kernel/kcov\_dataflow.c})
        and hardened remote-kworker collection path---including bounded
        per-section scratch sizing (\texttt{KCOV\_DF\_REMOTE\_ENABLE\_ARG}), an
        atomic input-epoch fence (\texttt{KCOV\_DF\_REMOTE\_RESET}), per-CPU drop
        counters, a page-unit shrinker, and a 64\,MiB idle pool cap in
        \texttt{kernel/kcov.c}---that rejects interrupt and nofault contexts and
        never interferes with existing KCOV infrastructure.
  \item A post-compilation IR pipeline enabling Rust module instrumentation
        without modifying \texttt{rustc}, plus native instrumentation via a
        custom-LLVM \texttt{rustc} that naturally preserves Rust's
        \texttt{dereferenceable(N)} contracts on \texttt{\&T}/\texttt{\&mut T}.
  \item A self-checking in-tree selftest suite---five kernel modules (three C
        incl.\ \texttt{pointee\_access\_c}, two Rust) and two user-space
        tests---alongside an IR-level pointee visibility audit across Linux
        x86-64, AArch64, and Rust kernel builds.
\end{itemize}

\section{Background \& Motivation}\label{sec:background}

\subsection{KCOV and Coverage-Guided Kernel Fuzzing}

KCOV~\cite{vyukov2016kcov} instruments the kernel at compile time via
\sancov's \tracepc callback. Each task owns a kernel-allocated buffer,
memory-mapped into user space, into which the kernel writes the program
counter of every executed edge. User-space fuzzers (primarily syzkaller)
hash these PCs into a coverage signal, retaining inputs that discover
novel edges in the corpus.

This architecture has proven remarkably effective: syzkaller continuously
discovers 100+ bugs per month across upstream kernels. However, coverage
\emph{saturates} on stateful subsystems. After initial exploration, new edges
become rare, and the fuzzer degenerates into random mutation without meaningful
guidance.

\subsection{The Data-Flow Blind Spot}

\begin{figure}[t]
\centering
\begin{tikzpicture}[
  node distance=0.3cm,
  box/.style={rectangle, draw, thick, minimum width=2.2cm, minimum height=0.6cm, font=\scriptsize\bfseries},
  msg/.style={-{Stealth[length=2mm]}, thick},
  lbl/.style={font=\scriptsize, midway, above},
  note/.style={font=\scriptsize\itshape, text=gray}
]
\node[box] (client) {Client (Fuzzer)};
\node[box, right=5cm of client] (server) {ksmbd (Kernel)};

\draw[dashed, gray] (client.south) -- ++(0,-5.5);
\draw[dashed, gray] (server.south) -- ++(0,-5.5);

\draw[msg] ([yshift=-0.8cm]client.south) -- node[lbl] {SMB2\_NEGOTIATE} ([yshift=-0.8cm]server.south);
\draw[msg] ([yshift=-1.4cm]server.south) -- node[lbl] {dialect OK} ([yshift=-1.4cm]client.south);

\draw[msg] ([yshift=-2.0cm]client.south) -- node[lbl] {SMB2\_SESSION\_SETUP} ([yshift=-2.0cm]server.south);
\draw[msg] ([yshift=-2.6cm]server.south) -- node[lbl] {authenticated} ([yshift=-2.6cm]client.south);

\draw[msg] ([yshift=-3.2cm]client.south) -- node[lbl] {SMB2\_TREE\_CONNECT} ([yshift=-3.2cm]server.south);
\draw[msg] ([yshift=-3.8cm]server.south) -- node[lbl] {share bound} ([yshift=-3.8cm]client.south);

\draw[msg, thick, red!70!black] ([yshift=-4.6cm]client.south) -- node[lbl, text=red!70!black] {SMB2\_CREATE(\colorbox{yellow!30}{DesiredAccess})} ([yshift=-4.6cm]server.south);
\draw[msg] ([yshift=-5.2cm]server.south) -- node[lbl] {file handle} ([yshift=-5.2cm]client.south);

\node[note, anchor=west] at ([xshift=0.5cm, yshift=-4.6cm]server.south) {$\to$ smb2\_open(work)};
\node[note, anchor=east] at ([xshift=-0.3cm, yshift=-1.1cm]client.south) {coverage saturates here};
\draw[decorate, decoration={brace, amplitude=3pt, mirror, raise=2pt}, gray]
  ([xshift=0.2cm, yshift=-0.6cm]client.south) -- ([xshift=0.2cm, yshift=-3.9cm]client.south)
  node[midway, left=5pt, note] {same edges};
\end{tikzpicture}
\caption{SMB2 protocol stages required before reaching the file-create path.
Edge coverage saturates after discovering the handshake sequence; the
security-relevant \texttt{DesiredAccess} field inside SMB2\_CREATE is
invisible to control-flow feedback.}\label{fig:ksmbd}
\end{figure}

Consider the \texttt{ksmbd} SMB2\_CREATE handler. Before a file operation
reaches the VFS layer, ksmbd enforces a multi-stage protocol handshake
between the client (fuzzer) and the in-kernel SMB server
(Figure~\ref{fig:ksmbd}). A fuzzer must produce valid negotiate, authenticate,
and tree-connect messages before it can exercise the file-create path. Once
coverage of this multi-stage sequence saturates, all subsequent CREATE requests
traverse identical edges regardless of \texttt{DesiredAccess}---the fuzzer
cannot distinguish \texttt{FILE\_READ\_DATA} from
\texttt{FILE\_WRITE\_DATA|FILE\_DELETE\_ON\_CLOSE}. With \toolname{}, the
\texttt{DesiredAccess} value becomes visible---though not, as it happens, at
the \texttt{smb2\_open()} boundary. That function takes a single
\texttt{struct ksmbd\_work *}, and the request lives two levels down, behind
that struct's \texttt{void *request\_buf}: an opaque pointer, whose pointee
layout the compiler cannot describe, followed by the request struct itself.
One level of dereference reaches \texttt{request\_buf}'s \emph{value} and stops
(Section~\ref{sec:deref}). The field is instead captured where a callee
receives it as a value---\texttt{smb\_map\_generic\_desired\_access(\_\_le32
daccess)}, and \texttt{smb\_check\_perm\_dacl()}, which takes
\texttt{req->DesiredAccess} as its fourth argument. Both are on the CREATE
path, so the value is recorded on every request; it is simply recorded one
frame deeper than the handler. This is the motivating example and also the
clearest case of why the one-level rule matters in practice.

Reaching those boundaries requires the remote path rather than plain per-task
collection: ksmbd processes requests on kernel work queues, not in the syscall
context of the client, so \texttt{fs/smb/server/} annotates its connection
handler and request dispatcher with
\texttt{kcov\_df\_remote\_start\_common()} /
\texttt{kcov\_df\_remote\_stop()} (Section~\ref{sec:device}). That is the
reason the transport cannot be per-task only.

A second example arises at Rust-to-C FFI boundaries. A Rust wrapper calls a C
function with a struct pointer whose fields determine kernel behavior. Edge
coverage sees one call site; the struct contents---which control whether a
privilege check is bypassed---are invisible.

\begin{figure}[t]
\centering
\begin{tikzpicture}[
  node distance=0.4cm and 1.2cm,
  block/.style={rectangle, draw, rounded corners, minimum width=2.2cm, minimum height=0.7cm, font=\footnotesize, align=center},
  arrow/.style={-{Stealth[length=2mm]}, thick},
  label/.style={font=\scriptsize\itshape, text=gray}
]
\node[font=\footnotesize\bfseries] (t1) {Edge Coverage (KCOV)};
\node[block, below=0.3cm of t1] (a1) {path\_openat()};
\node[block, below=of a1] (b1) {do\_open()};
\node[block, below=of b1] (c1) {vfs\_open\_consume()};
\draw[arrow] (a1) -- (b1);
\draw[arrow] (b1) -- (c1);
\node[label, right=0.1cm of b1] {op=?};
\node[label, below=0.1cm of c1] {Sees: A$\to$B$\to$C};

\node[font=\footnotesize\bfseries, right=3.5cm of t1] (t2) {+ Data-Flow Extraction};
\node[block, below=0.3cm of t2] (a2) {path\_openat()};
\node[block, below=of a2, fill=green!10] (b2) {do\_open()};
\node[block, below=of b2] (c2) {vfs\_open\_consume()};
\draw[arrow] (a2) -- (b2);
\draw[arrow] (b2) -- (c2);
\node[label, right=0.1cm of b2, text=black] {op$\to$\{0x8201,0,0x2,..\}};
\node[label, below=0.1cm of c2] {Sees: A$\to$B(open\_flag=0x8201)$\to$C};
\end{tikzpicture}
\caption{Edge coverage reports identical paths for different argument values.
Data-flow extraction reveals the security-relevant difference: in the v4
prototype (Figure~\ref{fig:calltree}), \texttt{do\_open()} expands
\texttt{const struct open\_flags *} at entry; under the v5 IR-only contract,
plain C pointers report addresses at entry while member loads/stores are
recorded via \texttt{0xA} pointee accesses.}\label{fig:blind-spot}
\end{figure}

\begin{figure}[t]
\centering
\begin{tikzpicture}[
  node distance=0.15cm,
  rec/.style={font=\ttfamily\tiny, anchor=west},
  entry/.style={rec, text=black!85},
  ret/.style={rec, text=black!60},
]
\node[entry] (l1) at (0,0)     {ENTRY do\_sys\_openat2\ \ arg[0] sz=4\ \ \{0xffffff9c\}\quad\textrm{\itshape AT\_FDCWD}};
\node[entry] (l2) at (0,-0.34) {ENTRY do\_sys\_openat2\ \ arg[1] sz=8\ \ \{0x7fff3eeaceb8\}\quad\textrm{\itshape user pathname}};
\node[entry] (l3) at (0,-0.68) {ENTRY do\_sys\_openat2\ \ arg[2] sz=24 @0xffff888008797ea0 \colorbox{yellow!30}{\{0x8201, 0, 0\}}};
\node[entry] (l4) at (0.4,-1.02) {ENTRY build\_open\_flags arg[0] sz=24 @0xffff888008797ea0 \colorbox{yellow!30}{\{0x8201, 0, 0\}}};
\node[entry] (l5) at (0.4,-1.36) {ENTRY build\_open\_flags arg[1] sz=20 \colorbox{orange!25}{@0xffff888008797e54} \{0, 0, 0, 0, 0\}};
\node[ret]   (l6) at (0.4,-1.70) {RET\ \ build\_open\_flags\ \ \ \ \ \ \ \ sz=4\ \ \{0\}};
\node[entry] (l7) at (0.4,-2.04) {ENTRY getname\_flags\ \ \ \ arg[0] sz=8\ \ \{0x7fff3eeaceb8\}};
\node[ret]   (l8) at (0.4,-2.38) {RET\ \ getname\_flags\ \ \ \ \ \ \ \ \ \ \ sz=192 @0xffff888008a847c0 \{0x..7d8, \colorbox{green!25}{0x2d66642f706d742f}\}};
\node[entry] (l9) at (0.4,-2.72) {ENTRY path\_openat\ \ \ \ \ \ arg[1] sz=20 \colorbox{orange!25}{@0xffff888008797e54} \{0x8201, 0, 0x2, 0x10000, 0x1\}};
\node[ret]   (la) at (0.8,-3.06) {RET\ \ vfs\_open\ \ \ \ \ \ \ \ \ \ \ \ \ \ \ sz=4\ \ \{0\}};
\node[ret]   (lb) at (0.4,-3.40) {RET\ \ path\_openat\ \ \ \ \ \ \ \ \ \ \ \ \ sz=255 @0xffff8880063044c0 \{...\}};
\node[ret]   (lc) at (0,-3.74)   {RET\ \ do\_sys\_openat2\ \ \ \ \ \ \ \ \ sz=4\ \ \colorbox{green!25}{\{0x4\}}\quad\textrm{\itshape the returned fd}};

\node[font=\tiny\itshape, anchor=west, text=black!65] at (0,-4.2)
  {\colorbox{yellow!30}{0x8201} = O\_WRONLY$|$O\_TRUNC$|$O\_LARGEFILE\quad
   \colorbox{green!25}{0x2d66642f706d742f} = \texttt{"/tmp/df-"}};
\node[font=\tiny\itshape, anchor=west, text=black!65] at (0,-4.5)
  {\colorbox{orange!25}{same address} at two boundaries: the \texttt{struct open\_flags}
   output parameter, empty at entry and filled later};
\end{tikzpicture}
\caption{An illustrative capture (from the v4 prototype) of one
\texttt{open("/tmp/df-target", O\_WRONLY|O\_TRUNC)}, excerpted from 8{,}881
records, with PCs resolved through \texttt{/proc/kallsyms}. Two things are
visible that edge coverage cannot show: the requested flags reaching the VFS as
\texttt{open\_how.flags}, and the same \texttt{struct open\_flags} object
observed empty at \texttt{build\_open\_flags} entry and populated by the time
\texttt{path\_openat} receives it. Under the v5 IR-only contract
(Section~\ref{sec:llvm-pass}), pointee bytes are snapshotted automatically for
\texttt{byval}/\texttt{sret}/\texttt{dereferenceable} arguments and captured on
actual loads/stores under \texttt{CONFIG\_KCOV\_DATAFLOW\_POINTEE\_ACCESS}.}\label{fig:calltree}
\end{figure}

\subsection{Threat Model and Assumptions}

I assume a standard kernel fuzzing threat model: an unprivileged user-space
process with access to the syscall interface attempts to trigger memory
corruption in kernel subsystems. KASAN/KMSAN detect the corruption once
triggered; my tool provides the \emph{path to triggering} it and the
\emph{data for triaging} it. Compile-time instrumentation is strictly
independent of debug metadata: building with or without \texttt{-g} emits the
identical sequence of callbacks, operands, and machine instructions. DWARF
metadata (\texttt{CONFIG\_DEBUG\_INFO}) is required only if an offline analyst or
tool (\texttt{sancov -symbolize-dataflow}, \texttt{addr2line}) wishes to map
recorded PCs and byte offsets back to source-level parameter and struct-member
names. The Kconfig options depend on \texttt{CONFIG\_KCOV},
\texttt{CC\_IS\_CLANG}, and \texttt{cc-option}/\texttt{rustc-option} probes for
the instrumentation flags, so they remain hidden on toolchains lacking the pass.

\section{System Design \& Architecture}\label{sec:design}

\subsection{Overview}

\toolname{} operates as a three-layer pipeline: (1)~compile-time
instrumentation inserts data-flow callbacks into kernel object files,
(2)~a runtime transport layer delivers records to user-space via a lock-free
per-task record buffer, and (3)~user-space consumers (fuzzer executors or analyst
tools) read the buffer via \texttt{mmap()} (Figure~\ref{fig:arch}). The system is activated by the
compiler flags \texttt{-fsanitize-coverage=trace-args,trace-ret} and
requires no source-level annotation.

\begin{figure}[t]
\centering
\begin{tikzpicture}[
  node distance=0.5cm and 0.8cm,
  layer/.style={rectangle, draw, thick, rounded corners, minimum width=9cm, minimum height=1.6cm, font=\footnotesize, align=center},
  comp/.style={rectangle, draw, fill=blue!8, minimum width=2.5cm, minimum height=0.6cm, font=\scriptsize, align=center},
  arrow/.style={-{Stealth[length=2mm]}, thick},
  label/.style={font=\scriptsize\itshape}
]
\node[layer, fill=yellow!10] (l1) {};
\node[font=\scriptsize\bfseries, anchor=north west] at ([xshift=3pt,yshift=-3pt]l1.north west) {Compile Time};
\node[comp] at ([yshift=-0.2cm]l1.center) (pass) {LLVM SanCov Pass\\(trace-args/ret)};

\node[layer, fill=orange!10, below=0.5cm of l1] (l2) {};
\node[font=\scriptsize\bfseries, anchor=north west] at ([xshift=3pt,yshift=-3pt]l2.north west) {Kernel Runtime};
\node[comp, fill=orange!15] at ([xshift=-2cm,yshift=-0.2cm]l2.center) (cb) {\_\_sanitizer\_cov\\\_trace\_args/ret};
\node[comp, fill=orange!15] at ([xshift=2cm,yshift=-0.2cm]l2.center) (buf) {Per-Task\\Record Buffer};

\node[layer, fill=green!10, below=0.5cm of l2] (l3) {};
\node[font=\scriptsize\bfseries, anchor=north west] at ([xshift=3pt,yshift=-3pt]l3.north west) {User Space};
\node[comp, fill=green!15] at ([yshift=-0.2cm]l3.center) (user) {mmap() consumer\\(fuzzer / analyst)};

\draw[arrow] (pass) -- node[left, label, pos=0.35] {inserts callbacks} (cb);
\draw[arrow] (cb) -- node[above, label] {reserve + write} (buf);
\draw[arrow] (buf) -- node[left, label, pos=0.35] {shared memory} (user);
\end{tikzpicture}
\caption{Three-layer architecture: compile-time instrumentation, kernel-resident
lock-free buffer, and user-space consumption via \texttt{mmap()}.}\label{fig:arch}
\end{figure}

\subsection{LLVM IR Boundary Instrumentation}\label{sec:llvm-pass}

The instrumentation pass resides in LLVM's \sancov{} framework
(\texttt{SanitizerCoverage.cpp}). Following review of upstream LLVM
proposals, v5 adopts a strict \emph{IR-only design rule}: debug information
may enrich offline symbolization, but must \emph{never} determine what
instrumentation is emitted. Building with \texttt{-g} emits the exact same
callbacks, operands, and machine code as building without \texttt{-g}; debug
metadata provides source locations for the calls and nothing else.

\paragraph{Callback ABI.} For every remaining IR function~$F$ after
inlining, the pass emits three-operand calls (Listing~\ref{lst:abi},
Figure~\ref{fig:abi}):
\begin{itemize}[nosep]
  \item At function entry (immediately before non-boundary edge/guard
        instrumentation):
        \texttt{\_\_sanitizer\_cov\_trace\_args(u64 tag, u64 val, const u64 *offsets)},
        where $\mathit{tag} = (\mathit{arg\_idx} \ll 32) \mid \mathit{size}$
        and $\mathit{arg\_idx}$ is the zero-based position in $F$'s IR
        parameter list.
  \item Before each non-\texttt{musttail} \texttt{ReturnInst}:
        \texttt{\_\_sanitizer\_cov\_trace\_ret(u64 size, u64 val, const u64 *offsets)}.
\end{itemize}
All calls are marked \texttt{notail} and \texttt{nomerge} so each observation
carries a unique return address, matching \texttt{trace-cmp}. The compiler
creates no \texttt{alloca}s, introduces no stack spills, and freezes potentially
\texttt{poison} values before passing them so \texttt{val} always holds bits.

\paragraph{Values, Objects, and IR-Verified Pointee Snapshots.}
The \texttt{offsets} argument dictates whether the runtime only logs
\texttt{val} or reads memory behind it:
\begin{enumerate}[nosep]
  \item \textbf{Values (\texttt{offsets == nullptr}):} \texttt{val} \emph{is}
        the value (zero-extended, low \texttt{size} bytes). Integers, pointers
        (recorded as the address they hold), floats, and vectors take this path;
        first-class aggregates passed or returned in registers are recursively
        extracted and emitted element by element with the same
        \texttt{arg\_idx}.
  \item \textbf{ABI-in-memory objects (\texttt{offsets != nullptr}):}
        arguments carrying \texttt{byval}, \texttt{byref}, \texttt{inalloca},
        or \texttt{preallocated}, and indirect \texttt{sret} return buffers,
        carry an IR-known memory layout. The pass walks the IR type (including
        nested structs and splitting elements $>8$ bytes or opaque \texttt{[N x
        i8]} Rust buffers into 8-byte chunks) to emit a private constant table
        where \texttt{offsets[0] = N} is followed by $N$ \emph{(byte offset,
        byte size)} pairs ($\leq 8$ bytes each).
  \item \textbf{Pointee snapshots (\texttt{offsets != nullptr}):} for a
        \texttt{noundef} pointer with IR attribute \texttt{dereferenceable(N)}
        or \texttt{dereferenceable\_or\_null(N)} (on entry, or on function
        return), \texttt{val} is the pointer and the pass emits an 8-byte-chunk
        table covering $\min(N, \text{\texttt{-sanitizer-coverage-trace-pointee-bytes}})$
        raw bytes (default~64; 0 disables snapshots). Plain C pointers lack IR
        dereferenceability contracts and are therefore reported strictly as
        addresses.
\end{enumerate}
Arguments unused by~$F$ (\texttt{use\_empty()}) and values without a
fixed-size bit pattern (e.g., scalable vectors, \texttt{void} returns) emit
\texttt{size = 0} (and null \texttt{val} for unused memory objects so dead
argument elimination cannot pass an unreadable poison pointer).

\paragraph{IR Arguments vs.\ Source Parameters and Inlining.}
Because instrumentation depends only on IR, \texttt{arg\_idx} reflects the ABI's
IR signature rather than source parameter numbering: a struct split across two
registers becomes two IR arguments, an indirect \texttt{sret} pointer or
\texttt{swifterror} slot is skipped from observation while preserving
subsequent argument indices, and an inlined callee no longer exists as an IR
function of its own (enabling \texttt{-fno-inline} increases callee boundary
visibility when desired). Offline consumers such as \texttt{sancov
-symbolize-dataflow} combine the call PC with binary descriptors and DWARF to
map \texttt{arg\_idx} and raw snapshot byte ranges back to source parameters and
struct fields. The pass flags are \texttt{-sanitizer-coverage-trace-args},
\texttt{-sanitizer-coverage-trace-ret}, and
\texttt{-sanitizer-coverage-trace-pointee-bytes}, exposed via Clang's
\texttt{-fsanitize-coverage=trace-args,trace-ret}.

\begin{figure}[t]
\centering
\begin{lstlisting}[language=,numbers=none,columns=fullflexible,keepspaces=true,basicstyle=\ttfamily\tiny,caption={Emitted v5 IR for \texttt{byval}/\texttt{sret} and scalar arguments under \texttt{-fsanitize-coverage=trace-args,trace-ret}. Notice \texttt{offsets[0]=N} leading word, packed \texttt{tag=(arg\_idx<<32)|size}, and \texttt{sret} consuming IR slot~0 while leaving \texttt{seed} at \texttt{arg\_idx=1}.},label=lst:abi]
@__sancov_offsets_0 = private constant [5 x i64] [i64 2, i64 0, i64 4, i64 8, i64 8]
@__sancov_offsets_1 = private constant [7 x i64] [i64 3, i64 0, i64 8, i64 8, i64 8, i64 16, i64 8]

; void submit(struct S * byval(struct S) s, int flags, void *token)
call void @__sanitizer_cov_trace_args(i64 16,          i64 %s_addr, ptr @__sancov_offsets_0) ; arg 0, sz 16
call void @__sanitizer_cov_trace_args(i64 4294967300,  i64 %flags,  ptr null)                ; arg 1, sz 4
call void @__sanitizer_cov_trace_args(i64 8589934600,  i64 %tok,    ptr null)                ; arg 2, sz 8
call void @__sanitizer_cov_trace_ret (i64 0,           i64 0,       ptr null)                ; void ret

; struct Big make_big(int seed) -- sret occupies IR slot 0, seed is arg_idx 1
call void @__sanitizer_cov_trace_args(i64 4294967300,  i64 %seed,   ptr null)                ; arg 1, sz 4
call void @__sanitizer_cov_trace_ret (i64 24,          i64 %sret,   ptr @__sancov_offsets_1) ; 24B sret
\end{lstlisting}
\end{figure}

\begin{figure}[t]
\centering
\begin{tikzpicture}[
  node distance=0.35cm,
  sig/.style={rectangle, draw, thick, rounded corners, fill=blue!8, minimum width=9.0cm, minimum height=0.55cm, font=\ttfamily\scriptsize, align=center, inner sep=4pt},
  cell/.style={rectangle, draw, minimum width=1.15cm, minimum height=0.5cm, font=\tiny, align=center},
  tbl/.style={cell, fill=yellow!15},
  note/.style={font=\tiny\itshape, text=black!65},
]
\node[sig] (a) {trace\_args(\colorbox{orange!18}{tag = (arg\_idx << 32) | size}, \colorbox{green!18}{val}, offsets)};
\node[sig, below=0.3cm of a] (r) {trace\_ret(\colorbox{orange!18}{size}, \colorbox{green!18}{val}, offsets)};

\node[note, below=0.35cm of r.west, anchor=north west] (n1)
  {\texttt{offsets = null}: \texttt{val} \emph{is} the value (scalar, pointer addr, or scalar element); or \texttt{KCOV\_DF\_TYPE\_ACCESS} value};
\node[note, below=0.08cm of n1.west, anchor=north west] (n2)
  {\texttt{offsets != null}: \texttt{val} is an IR-verified address (\texttt{byval}/\texttt{sret}/\texttt{dereferenceable(N)}), \texttt{size} bytes total};

\node[tbl, fill=yellow!30, below=0.5cm of n2.west, anchor=north west] (tlen) {N};
\node[tbl, right=0pt of tlen] (t0) {off$_0$};
\node[tbl, right=0pt of t0] (t1) {sz$_0$};
\node[tbl, right=0pt of t1] (t2) {off$_1$};
\node[tbl, right=0pt of t2] (t3) {sz$_1$};
\node[cell, right=0pt of t3, draw=none] (t4) {\ldots};
\node[note, right=0.1cm of t4] {2N+1 words, \texttt{.rodata}};
\node[note, left=0.1cm of tlen] {offsets:};
\end{tikzpicture}
\caption{The v5 three-operand callback ABI. \texttt{trace\_args} packs IR
\texttt{arg\_idx} and \texttt{size} into a single 64-bit \texttt{tag};
when non-null, \texttt{offsets[0] = N} gives the pair count followed by
\emph{(offset, size)} pairs ($\leq 8$\,B each). Neither callback takes a
program counter.}\label{fig:abi}
\end{figure}

\paragraph{Pointer-Access Tracking (Follow-on Mode).}
As measured in Section~\ref{sec:eval}, ordinary C pointer arguments (\texttt{T *})
rarely carry IR \texttt{dereferenceable} attributes. On the companion
pointee-access extension (\texttt{-sanitizer-coverage-trace-pointee-access} in
LLVM paired with \texttt{CONFIG\_KCOV\_DATAFLOW\_POINTEE\_ACCESS} in Linux), the
pass emits \texttt{trace\_args} calls after actual loads and stores performed by
the function through a pointer argument at a constant offset, passing bit~31
(\texttt{KCOV\_DF\_ARG\_ACCESS}) in \texttt{arg\_idx}, the loaded or stored
register value as \texttt{val}, and \texttt{offsets = null}. Because zero
extra memory reads are added, KASAN and KCSAN observe identical memory accesses.

\subsection{Lock-Free Per-Task Record Buffer}\label{sec:ringbuf}

The buffer is append-only rather than circular, which is worth stating plainly
because ``ring'' would imply the opposite failure mode: records accumulate from
\texttt{area[1]} upward, \texttt{area[0]} counts the words committed so far,
and once a record no longer fits it is dropped. A trace that outgrows the
buffer therefore loses its \emph{tail}, not its head---the beginning of an
execution is preserved, which is what a consumer reconstructing a call sequence
needs. This matches mainline \texttt{kcov}'s buffer discipline.

The kernel backend adds per-task fields to \texttt{task\_struct}:
\texttt{kcov\_df\_area} (pointer to the \texttt{u64} buffer),
\texttt{kcov\_df\_size} (capacity in words), \texttt{kcov\_df\_enabled},
\texttt{kcov\_df\_seq} (sequence counter plus recursion guard), and, for the
remote path, \texttt{kcov\_df}, \texttt{kcov\_df\_remote\_depth}, and
\texttt{kcov\_df\_reset\_seq} (epoch snapshot for safe input resets).

\paragraph{Write Path.} The callback implementation uses
\texttt{READ\_ONCE}/\texttt{WRITE\_ONCE} to reserve a contiguous slot in
the buffer (portable to ARM64 without architecture-specific atomics), under the
same single-writer discipline as mainline \texttt{kcov.c}: the record length is
validated first and the word count at \texttt{area[0]} is committed only once
the record fits, so \texttt{area[0]} is always an exact high-water mark of
written words. If the record does not fit, \texttt{kcov\_df\_reserve()}
increments the per-CPU drop counter \texttt{kcov\_df\_dropped\_records}
(\texttt{this\_cpu\_inc()}) and drops the event without blocking or locking.
The callbacks take no program counter; the recorded PC is the callback's return
address minus one byte (\texttt{\_RET\_IP\_ - 1}), so it lands inside the call
instruction rather than on the instruction after it, matching unwinder and
\texttt{sancov -symbolize-dataflow} expectations. The PC is then canonicalized
by subtracting the KASLR offset, as mainline \kcov{} does.

When \texttt{offsets} is \texttt{NULL} the value arrived in a register and is
stored as it came---nothing is dereferenced. When \texttt{offsets} is
non-\texttt{NULL}, \texttt{val} is an address and \texttt{offsets[0]} holds the
pair count $N$ (read safely via \texttt{copy\_from\_kernel\_nofault()} and
clamped to \texttt{KCOV\_DF\_MAX\_VALS}); each pair is read as
$\mathit{field\_addr} = \mathit{val} + \mathit{offsets}[1 + 2i]$ for
$\mathit{offsets}[1 + 2i + 1]$ bytes, clamped to 8, via
\texttt{copy\_from\_kernel\_nofault()}. A \texttt{NULL} or
\texttt{IS\_ERR\_VALUE} address ($[-4095,-1]$), a field that cannot be read,
and an unused object or unreportable value sent with \texttt{size}~0 all store
the sentinel \texttt{KCOV\_DF\_MAGIC\_BAD} (\texttt{0xBADADD85}), so a missing
value is distinguishable from a genuine zero. The callbacks are bracketed by
\texttt{user\_access\_save()}/\texttt{restore()} so that a record written from
a function with SMAP/PAN user access enabled neither trips objtool nor leaves
user access open across the collector.

\paragraph{Record Format.} Each record is a variable-length run of
\texttt{u64} words: a three-word header followed by $\mathit{nvals}$ value
words, where $\mathit{nvals}$ is the count $N$ from \texttt{offsets[0]}, or~1
for scalar/pointer values and \texttt{ACCESS} records, or~2 for comparisons.
Word~[0] packs the sequence number (bits 0--23), \texttt{KCOV\_DF\_HDR\_MEM}
(bit~24, set whenever the value words were read from memory at word~[2]),
the record type (bits 28--31: \texttt{0xE} entry, \texttt{0xF} return,
\texttt{0xC} comparison, \texttt{0xA} pointer access), $\mathit{nvals}$
(bits 32--47), the value or access size in bytes (bits 48--55, clamped to 255),
and the IR argument index (bits 56--63, clamped to 255; zero for a return).
Word~[1] is the canonicalized PC. For \texttt{ENTRY}/\texttt{RET}, word~[2] is
the address the bytes were read from (with \texttt{KCOV\_DF\_HDR\_MEM} set) or
zero; for \texttt{CMP}, it carries \texttt{KCOV\_CMP\_*} flags; and for
\texttt{KCOV\_DF\_TYPE\_ACCESS} (\texttt{0xA}), it packs
\texttt{KCOV\_DF\_ACCESS\_STORE}, \texttt{\_VOLATILE}, \texttt{\_ATOMIC}, and
LLVM atomic ordering bits while word~[3] holds the loaded or stored value with
\texttt{KCOV\_DF\_HDR\_MEM} cleared (Figure~\ref{fig:record}).
The \texttt{offsets} tables remain in \texttt{.rodata} and never reach user
space.

\paragraph{Unified stream.} When \texttt{CONFIG\_KCOV\_ENABLE\_COMPARISONS} is
also set, the shared \texttt{\_\_sanitizer\_cov\_trace\_cmp*()} callbacks fan
out into the same buffer as \texttt{0xC} records carrying both operands and the
\texttt{KCOV\_CMP\_SIZE()}/\texttt{KCOV\_CMP\_CONST} type bits, so
input-to-state operand pairs and boundary arguments arrive interleaved in one
sequence-numbered stream rather than in two buffers that must be correlated
afterwards. An older consumer walking records by \texttt{pos += 3 + nvals}
skips unknown record types such as \texttt{0xA} safely without losing framing.

\begin{figure}[t]
\centering
\begin{tikzpicture}[
  cell/.style={rectangle, draw, minimum width=1.45cm, minimum height=0.52cm, font=\tiny, align=center},
  hdr/.style={cell, fill=blue!12},
  dat/.style={cell, fill=green!12},
  cnt/.style={cell, fill=gray!15},
  bit/.style={rectangle, draw, minimum height=0.45cm, font=\tiny, align=center, fill=blue!6},
]
\node[cnt] (c0) {area[0]\\word count};
\node[hdr, right=0pt of c0] (h1) {[0] header};
\node[hdr, right=0pt of h1] (h2) {[1] pc};
\node[hdr, right=0pt of h2] (h3) {[2] addr / cmp /\\access kind};
\node[dat, right=0pt of h3] (d1) {val$_0$};
\node[dat, right=0pt of d1] (d2) {val$_1$};
\node[cell, right=0pt of d2, draw=none] (dots) {\ldots};

\node[font=\tiny, below=0.15cm of c0] {u64};
\node[font=\tiny\itshape, above=0.15cm of h2] {3-word header};
\node[font=\tiny\itshape, above=0.15cm of d1] {nvals words};

\draw[decorate, decoration={brace, amplitude=3pt, raise=2pt}]
  (h1.north west) -- (h3.north east);
\draw[decorate, decoration={brace, amplitude=3pt, raise=2pt}]
  (d1.north west) -- (d2.north east);

\node[bit, minimum width=1.4cm, below=0.88cm of c0.west, anchor=north west] (b0) {seq\\{}[23:0]};
\node[bit, minimum width=0.85cm, right=0pt of b0, fill=orange!15] (bm) {MEM\\{}[24]};
\node[bit, minimum width=0.95cm, right=0pt of bm] (b1) {type\\{}[31:28]};
\node[bit, minimum width=1.15cm, right=0pt of b1] (b2) {nvals\\{}[47:32]};
\node[bit, minimum width=1.05cm, right=0pt of b2] (b3) {size\\{}[55:48]};
\node[bit, minimum width=1.25cm, right=0pt of b3] (b4) {arg\_idx\\{}[63:56]};
\node[font=\tiny\itshape, right=0.15cm of b4] {word [0]};
\end{tikzpicture}
\caption{Buffer layout: \texttt{area[0]} holds the committed word count; each
record is a three-word header followed by \emph{nvals} value words. Bit~24
(\texttt{KCOV\_DF\_HDR\_MEM}) marks memory-read snapshots; word~[2] holds the
address (when MEM=1), comparison flags (\texttt{0xC}), or access kind
(\texttt{0xA}).}\label{fig:record}
\end{figure}

\paragraph{Safety.} Callbacks are marked \texttt{notrace},
\texttt{\_\_no\_sanitize\_coverage}, and \texttt{noinline} to prevent recursion,
and bit~31 of \texttt{kcov\_df\_seq} is a per-task guard held for the duration
of a record so that instrumentation reached from inside the collector cannot
re-enter it. No \texttt{printk} or allocation occurs in the data path. The write
path bails out when \texttt{!in\_task() || pagefault\_disabled()}: the first
rejects softirq, hardirq, and NMI, and the second contains the
self-instrumentation storm that otherwise results when an instrumented leaf is
reached from inside a \texttt{copy\_from\_kernel\_nofault()} region---a single
KASAN stack walk would otherwise issue a callback per load.

\subsection{Separate Device Interface}\label{sec:device}

\toolname{} exposes \texttt{/sys/kernel/debug/kcov\_dataflow} as a device
\emph{completely independent} from the existing \texttt{/sys/kernel/debug/kcov}.
This ensures zero interference with syzkaller's existing KCOV usage.

The ioctl namespace uses magic~\texttt{'d'}:
\begin{itemize}[nosep]
  \item \texttt{KCOV\_DF\_INIT\_TRACK} (\texttt{\_IOR('d', 1, unsigned long)}):
        set the buffer size in \texttt{u64} words (up to 128\,MiB), by value,
        as \texttt{KCOV\_INIT\_TRACE} does
  \item \texttt{KCOV\_DF\_ENABLE} (\texttt{\_IO('d', 100)}): start recording
        into the calling task's own buffer (refuses with \texttt{-EBUSY} if a
        remote handle is published or remote workers remain in flight)
  \item \texttt{KCOV\_DF\_DISABLE} (\texttt{\_IO('d', 101)}): stop recording
  \item \texttt{KCOV\_DF\_REMOTE\_ENABLE} (\texttt{\_IOW('d', 102, \_\_u64)}):
        publish a \texttt{kcov\_remote\_handle()}-encoded handle so that kernel
        threads and other tasks annotated with
        \texttt{kcov\_df\_remote\_start()} record into this buffer; passed by
        pointer so the full 64-bit handle survives 32-bit and compat callers
  \item \texttt{KCOV\_DF\_REMOTE\_DISABLE} (\texttt{\_IO('d', 103)}): withdraw
        the handle
  \item \texttt{KCOV\_DF\_REMOTE\_ENABLE\_ARG} (\texttt{\_IOW('d', 104, struct
        kcov\_df\_remote\_arg)}): publish a remote handle together with an
        explicit per-section scratch size \texttt{area\_size} (in \texttt{u64}
        words, bounded to $[8, 2^{20}]$ and $\leq \mathit{df\to size}$; zero
        selects the default clamp) so high-concurrency subsystems with short
        work items do not allocate 8\,MiB per concurrent kworker
  \item \texttt{KCOV\_DF\_REMOTE\_RESET} (\texttt{\_IO('d', 105)}): atomically
        bump the buffer's epoch counter (\texttt{reset\_seq}), execute
        \texttt{smp\_wmb()}, and zero \texttt{area[0]}. Remote sections that
        started under an earlier epoch drop their scratch batch at merge
        instead of bleeding stale records into the next fuzzing input, which a
        plain \texttt{mmap} write to \texttt{area[0]} cannot prevent.
\end{itemize}
User-space maps the buffer via \texttt{mmap()} for zero-copy reads and walks it
as \texttt{pos += KCOV\_DF\_RECORD\_WORDS(nvals)} while
\texttt{pos < 1 + area[0]}. Remote writers assemble records in a private
per-worker scratch buffer and publish them into the shared area with a single
\texttt{atomic64\_try\_cmpxchg()} cursor update, so many kworkers can merge
without a lock.

\paragraph{Bounded Remote Pools and Memory Reclamation.}
Deployment on high-concurrency kernel servers (such as \texttt{ksmbd} workers
blocked in oplock waits) exposes a hazard shared by mainline \texttt{kcov} and
\toolname{}: parking every remote scratch buffer forever ratchets idle
\texttt{vmalloc} memory up to peak concurrency until the host panics with OOM.
Accordingly, v5 introduces three defenses across both subsystems: (1)~per-section
\texttt{kcov\_df\_remote\_arg::area\_size} sizing and best-effort scratch
allocation using \texttt{GFP\_KERNEL | \_\_GFP\_NORETRY | \_\_GFP\_NOWARN}
(counting failed or truncated merges per CPU in read-only debugfs files
\texttt{kcov\_dataflow\_dropped\_records} and
\texttt{kcov\_dataflow\_dropped\_sections}, while keeping inert-context skips
uncounted so global \texttt{INSTRUMENT\_ALL} overhead stays zero); (2)~a
registered kernel shrinker (\texttt{kcov-df-scratch}) that tracks idle pooled
memory in pages (\texttt{kcov\_df\_scratch\_pool\_pages}) and reclaims whole
scratch areas under memory pressure using \texttt{mutex\_trylock} to avoid lock
inversion; and (3)~in mainline \texttt{kernel/kcov.c}, a 64\,MiB cap
(\texttt{KCOV\_REMOTE\_POOL\_MAX\_BYTES}) on idle task-context remote areas
together with \texttt{vfree\_atomic()} and \texttt{\_\_GFP\_NORETRY |
\_\_GFP\_NOWARN} allocation.

\subsection{Rust Module Support}\label{sec:rust}

The Rust compiler (\texttt{rustc}) bundles its own LLVM (v23 for
rustc~1.101), which lacks our new callbacks. Two instrumentation paths exist:

\paragraph{Path A: Post-compilation pipeline (no rustc modification).}
For environments using an unmodified rustc:
\begin{enumerate}[nosep]
  \item \texttt{rustc --emit=llvm-ir}: emit LLVM IR text
  \item \texttt{opt -passes=sancov-module -sanitizer-coverage-level=1} \\
        \texttt{-sanitizer-coverage-trace-args -sanitizer-coverage-trace-ret}:
        instrument with our pass
  \item \texttt{llc -filetype=obj}: compile to object code
  \item Standard module linking produces the instrumented \texttt{.ko}
\end{enumerate}
The boundary flags require an active SanitizerCoverage pass run (the Clang
driver supplies the default \texttt{edge} level automatically whenever any
\texttt{-fsanitize-coverage=} mode is given, while emitting zero block or edge
callbacks unless a block-level mode such as \texttt{trace-pc} is requested).
Because the pass operates solely on LLVM IR, it works uniformly on C and Rust
IR (Figure~\ref{fig:rust-pipeline}), and \texttt{rustc} naturally annotates
safe Rust references (\texttt{\&T}, \texttt{\&mut T}) with \texttt{noundef
dereferenceable(size\_of::<T>())} and \texttt{Option<\&T>} with
\texttt{dereferenceable\_or\_null}, yielding automatic entry pointee snapshots
without any DWARF metadata. \texttt{rustc} lowers memory-passed Rust structs
to opaque \texttt{[N x i8]} IR types, which \toolname{} splits into 8-byte
words up to the pointee budget.

\begin{figure}[t]
\centering
\begin{tikzpicture}[
  node distance=0.3cm and 0.4cm,
  pstep/.style={rectangle, draw, fill=blue!8, minimum width=1.8cm, minimum height=0.6cm, font=\scriptsize, align=center},
  arrow/.style={-{Stealth[length=2mm]}, thick},
  label/.style={font=\tiny\itshape, text=gray}
]
\node[pstep] (rustc) {rustc\\--emit=llvm-ir};
\node[pstep, right=0.6cm of rustc] (opt) {opt\\(trace-args/ret)};
\node[pstep, right=0.6cm of opt] (llc) {llc\\-filetype=obj};
\node[pstep, right=0.6cm of llc] (ld) {ld\\module.ko};

\draw[arrow] (rustc) -- node[above, label] {.ll} (opt);
\draw[arrow] (opt) -- node[above, label] {.ll} (llc);
\draw[arrow] (llc) -- node[above, label] {.o} (ld);

\node[label, below=0.15cm of rustc] {Rust LLVM 23};
\node[label, below=0.15cm of opt] {Custom LLVM 24};
\node[label, below=0.15cm of llc] {Custom LLVM 24};
\end{tikzpicture}
\caption{Rust post-compilation pipeline: text-format IR avoids bitcode version
incompatibility between rustc's bundled LLVM and our pass.}\label{fig:rust-pipeline}
\end{figure}

\paragraph{Path B: Native instrumentation (custom rustc).}
Building \texttt{rustc} against the custom LLVM~24 (via
\texttt{llvm-config} in \texttt{config.toml}) gives rustc native support for
\texttt{-Cllvm-args=-sanitizer-coverage-trace-args/ret} (alongside a coverage
level when invoking LLVM directly). The kernel build system passes these flags
transparently via \texttt{RUSTFLAGS\_KCOV\_DATAFLOW}---no post-compilation step
required. \textbf{This is the required path for tracing Rust kernel modules}, as
Kbuild compiles \texttt{core}, \texttt{alloc}, and \texttt{kernel} crates
internally and does not expose intermediate IR for external instrumentation.

\subsection{Pointer Dereference, Snapshots, and Access Records}\label{sec:deref}

A key design question is how \toolname{} captures data behind pointer
arguments (heap, stack, or static memory) safely without guessing pointer
validity.

\paragraph{IR-Bounded Shallow Snapshots (One Level).} Whenever the compiler
emits a non-null \texttt{offsets} table---either for an ABI-in-memory object
(\texttt{byval}/\texttt{sret}) or for a \texttt{noundef dereferenceable(N)} /
\texttt{dereferenceable\_or\_null(N)} pointer---the kernel callback performs
\emph{one level of pointer dereference} at the trace point via
\texttt{copy\_from\_kernel\_nofault()}, up to
\texttt{-sanitizer-coverage-trace-pointee-bytes} (default 64) and clamped to
\texttt{KCOV\_DF\_MAX\_VALS}:
\begin{itemize}[nosep]
  \item \textbf{Freed slab (KASAN poison):} Safely reads poison bytes
        (\texttt{0xfd}/\texttt{0xff})---revealing the object was freed.
  \item \textbf{Unmapped page:} \texttt{copy\_from\_kernel\_nofault()} returns
        error; the word is set to \texttt{KCOV\_DF\_MAGIC\_BAD}.
  \item \textbf{ERR\_PTR or NULL:} Checked via \texttt{IS\_ERR\_VALUE} /
        null check; nothing is dereferenced and every value word is marked
        \texttt{KCOV\_DF\_MAGIC\_BAD}.
  \item \textbf{Fields or chunks wider than 8 bytes:} Split by the compiler's
        \texttt{offsets} table into 8-byte pieces (or clamped to 8 bytes per
        pair in the collector).
\end{itemize}
Because pointers stored \emph{inside} a snapshotted region are recorded as
addresses without chasing nested pointers, capture cost stays $O(n)$ in the
number of pairs and avoids recursive faults. Nested structs embedded \emph{by
value} inside a \texttt{byval}/\texttt{sret} IR type are traversed recursively
by \texttt{getObjectOffsets()}.

\paragraph{Closing the Plain-C Pointer Gap via \texttt{0xA} Access Records.}
Because C rarely equips ordinary \texttt{T *} parameters with IR
\texttt{dereferenceable(N)} contracts (Section~\ref{sec:eval}), entry snapshots
alone only observe the address held by a plain C pointer. With the optional
pointee-access extension (\texttt{CONFIG\_KCOV\_DATAFLOW\_POINTEE\_ACCESS},
plus \texttt{\_VOLATILE} for \texttt{READ\_ONCE}/\texttt{WRITE\_ONCE} and
\texttt{\_ATOMIC} for compiler atomics), LLVM instruments loads and stores
through pointer arguments at constant byte offsets directly after the memory
operation completes, passing the already-loaded or stored register value as a
\texttt{KCOV\_DF\_TYPE\_ACCESS} (\texttt{0xA}) record. Unlike entry snapshots,
access records reflect the pointee's value at the moment of the access inside
the callee (not at entry) and execute \emph{zero} extra memory reads, keeping
KASAN and KCSAN behavior invariant.

\section{Implementation}\label{sec:impl}

\subsection{LLVM Pass (\texorpdfstring{$\sim$}{\~{}}350~LOC) and Upstream Stack}

The upstream-targeted stack spans four commits across LLVM, Clang, and
\texttt{compiler-rt}:
\begin{enumerate}[nosep]
  \item \textbf{LLVM Pass (\texttt{SanitizerCoverage.cpp}, $+$350~LOC):}
        adds \texttt{InjectTraceForArgs()} and \texttt{InjectTraceForRet()}
        together with \texttt{emitTraceCall()}, \texttt{traceValue()},
        \texttt{traceObject()}, \texttt{tracePointee()},
        \texttt{getPointeeBytes()}, and \texttt{getObjectOffsets()}. Every
        emitted \texttt{CallInst} sets \texttt{TCK\_NoTail} and
        \texttt{setCannotMerge()}. Arguments with \texttt{sret} or
        \texttt{swifterror} are skipped while keeping their IR positions.
        \texttt{FieldOffsetsCache} memoizes emitted \texttt{\_\_sancov\_offsets\_}
        constants (prefixed by pair count \texttt{N}) per IR type, and
        \texttt{tracePointee()} uses \texttt{ArrayType::get(i8, Bytes)} to cut
        \texttt{noundef dereferenceable} buffers into 8-byte words capped by
        \texttt{-sanitizer-coverage-trace-pointee-bytes} (default~64).
        \texttt{InjectTraceForArgs()} runs ahead of \texttt{InjectCoverage()}
        after static allocas, and \texttt{InjectTraceForRet()} runs before
        \texttt{InjectTraceForExits()}, with potentially \texttt{poison} values
        frozen before widening.
  \item \textbf{Clang Driver:} adds \texttt{SanitizeCoverageTraceArgs} and
        \texttt{SanitizeCoverageTraceRet} in \texttt{CodeGenOptions.def/.h},
        parses \texttt{trace-args} and \texttt{trace-ret} in
        \texttt{SanitizerArgs.cpp} and \texttt{Options.td} (forwarding via
        \texttt{-fsanitize-coverage-trace-args}/\texttt{-ret} cc1 flags), wires
        them in \texttt{BackendUtil.cpp}, and updates
        \texttt{hasSanitizeCoverage()} so specifying either flag alone runs the
        pass without emitting block/edge callbacks.
  \item \textbf{\texttt{compiler-rt} Weak Runtime:} declares both callbacks in
        \texttt{sanitizer\_interface\_internal.h} and
        \texttt{sanitizer\_coverage\_interface.inc}, and adds weak, empty default
        definitions in \texttt{sanitizer\_coverage\_libcdep\_new.cpp} so binaries
        built with \texttt{-fsanitize-coverage=trace-args,trace-ret} link
        without supplying a custom runtime.
  \item \textbf{Documentation:} adds the formal specification and IR-attribute
        contract to \texttt{clang/docs/SanitizerCoverage.md}.
\end{enumerate}
Follow-on companion branches house \texttt{\_\_sancov\_dataflow} descriptors
and \texttt{sancov -symbolize-dataflow} offline symbolization, the updated
3-operand libFuzzer \texttt{HandleDataflow()} consumer, and
\texttt{-sanitizer-coverage-trace-pointee-access} for \texttt{0xA} pointer
load/store instrumentation.

\subsection{Kernel Backend (\texorpdfstring{$\sim$}{\~{}}1{,}424~LOC)}

The kernel implementation in \texttt{kernel/kcov\_dataflow.c} (1{,}424~LOC,
separate from \texttt{kernel/kcov.c}) plus \texttt{include/uapi/linux/kcov\_dataflow.h}
adds:
\begin{itemize}[nosep]
  \item \texttt{struct kcov\_dataflow}: owns buffer, size, refcount, remote
        handle, \texttt{reset\_seq} epoch, and \texttt{file\_operations}.
  \item \texttt{\_\_sanitizer\_cov\_trace\_args(u64 tag, u64 val, const u64
        *offsets)}: unpacks $\mathit{arg\_idx}$ and $\mathit{size}$ from
        \texttt{tag}; when bit~31 (\texttt{KCOV\_DF\_ARG\_ACCESS}) is set,
        routes directly to \texttt{kcov\_df\_write(KCOV\_DF\_TYPE\_ACCESS,
        ...)} with \texttt{offsets = NULL} and word~[2] encoding access flags;
        otherwise emits \texttt{KCOV\_DF\_TYPE\_ENTRY}. Marked \texttt{notrace},
        \texttt{\_\_no\_sanitize\_coverage}, \texttt{noinline}, bracketed by
        \texttt{user\_access\_save()}/\texttt{restore()}, using
        \texttt{\_RET\_IP\_ - 1} for PC.
  \item \texttt{\_\_sanitizer\_cov\_trace\_ret(u64 size, u64 val, const u64
        *offsets)}: same guard pattern writing \texttt{KCOV\_DF\_TYPE\_RET}
        with argument index~0.
  \item \texttt{kcov\_df\_write()} and \texttt{kcov\_df\_reserve()}:
        context/recursion guards, \texttt{this\_cpu\_inc(kcov\_df\_dropped\_records)}
        on overflow, header packing (including \texttt{KCOV\_DF\_HDR\_MEM}),
        and either register storage or \texttt{copy\_from\_kernel\_nofault()}
        traversal of \texttt{offsets[0]=N}.
  \item \texttt{kcov\_df\_trace\_cmp()}: fan-out from
        \texttt{\_\_sanitizer\_cov\_trace\_cmp*()} behind static keys when
        \texttt{CONFIG\_KCOV\_ENABLE\_COMPARISONS} is enabled, writing
        \texttt{0xC} records into the same stream.
  \item \texttt{kcov\_df\_remote\_start()}/\texttt{stop()},
        \texttt{kcov\_df\_merge()}, and \texttt{kcov-df-scratch} shrinker:
        bounded per-worker scratch areas (\texttt{KCOV\_DF\_REMOTE\_ENABLE\_ARG},
        clamped to $[8, 2^{20}]$ words), atomic epoch filtering
        (\texttt{KCOV\_DF\_REMOTE\_RESET}), per-CPU section drop counters
        (\texttt{kcov\_df\_dropped\_sections}), and page-unit VM reclamation via
        \texttt{kcov\_df\_scratch\_pool\_pages}. Companion hardening in
        \texttt{kernel/kcov.c} adds a 64\,MiB cap
        (\texttt{KCOV\_REMOTE\_POOL\_MAX\_BYTES}) on idle task-context remote
        areas with \texttt{vfree\_atomic()} and \texttt{\_\_GFP\_NORETRY |
        \_\_GFP\_NOWARN}.
\end{itemize}

Both callbacks sit on objtool's \texttt{uaccess\_safe\_builtin[]} list,
matching \texttt{\_\_sanitizer\_cov\_trace\_cmp*()}.
Kconfig options \texttt{CONFIG\_KCOV\_DATAFLOW\_ARGS} and \texttt{\_RET} gate
core tracing; optional \texttt{CONFIG\_KCOV\_DATAFLOW\_POINTEE\_ACCESS} (with
\texttt{\_VOLATILE} and \texttt{\_ATOMIC}, all default \texttt{n}) gates
\texttt{0xA} instrumentation on capable compilers without changing the record
stream when disabled.

\subsection{Build System Integration}

\texttt{scripts/Makefile.kcov} exports \texttt{CFLAGS\_KCOV\_DATAFLOW},
\texttt{RUSTFLAGS\_KCOV\_DATAFLOW}, and \texttt{CFLAGS\_KCOV\_DATAFLOW\_NO\_INLINE}.
\texttt{scripts/Makefile.lib} applies them per object (\texttt{KCOV\_DATAFLOW\_my\_module.o := y})
or directory (\texttt{KCOV\_DATAFLOW := y}) uniformly across C and Rust.

\paragraph{Global Enablement and Optional Inlining Control.}
\texttt{CONFIG\_KCOV\_DATAFLOW\_INSTRUMENT\_ALL} mirrors
\texttt{CONFIG\_KCOV\_INSTRUMENT\_ALL}, with per-file or per-directory
\texttt{:= n} opt-outs. Under the IR-only design, inlined callees are folded into
their callers and no longer carry separate IR function boundaries; therefore,
enabling \texttt{CONFIG\_KCOV\_DATAFLOW\_NO\_INLINE} (\texttt{-fno-inline} on
explicitly opted-in modules) directly increases boundary visibility by keeping
small helpers as distinct calls---making PCs resolvable via \texttt{kallsyms}
and records one-to-one with source call sites. Because some kernel code relies
on inlining to fold constants into inline-asm \texttt{"i"} operands,
\texttt{NO\_INLINE} intentionally applies only to per-file opt-ins rather than
\texttt{INSTRUMENT\_ALL}.

\subsection{User-Space Runtime and Consumer}\label{sec:libfuzzer}

In user space, \texttt{compiler-rt}'s weak empty callbacks ensure binaries
compile and run even without an active consumer. Stronger definitions---such as
libFuzzer's \texttt{HandleDataflow()} (updated on the follow-on libFuzzer branch
to unpack \texttt{tag}, read the \texttt{offsets[0]=N} table, and skip
\texttt{KCOV\_DF\_ARG\_ACCESS} or fold values by call site, IR argument, and
element)---override the weak stubs at link time. Unlike kernel records, which
strip \texttt{.rodata} offset tables before writing to \texttt{area[]}, a
user-space runtime shares the address space and can inspect \texttt{offsets}
directly.

In the kernel workflow, a minimal consumer opens
\texttt{/sys/kernel/debug/kcov\_dataflow}, allocates with
\texttt{KCOV\_DF\_INIT\_TRACK}, \texttt{mmap()}s the buffer, uses either
\texttt{KCOV\_DF\_ENABLE} (local) or
\texttt{KCOV\_DF\_REMOTE\_ENABLE\_ARG} / \texttt{KCOV\_DF\_REMOTE\_RESET}
(remote kworkers), and walks records via
\texttt{pos += KCOV\_DF\_RECORD\_WORDS(KCOV\_DF\_HDR\_NVALS(hdr))}, checking
\texttt{KCOV\_DF\_HDR\_MEM} to distinguish dereferenced object/pointee snapshots
from register values and \texttt{KCOV\_DF\_TYPE\_ACCESS} for mid-call
loads/stores. Offline \texttt{sancov -symbolize-dataflow} or \texttt{addr2line}
maps recorded PCs and byte offsets to source-level parameter and field names.

\section{Evaluation}\label{sec:eval}

\subsection{Experimental Setup and v4 \texorpdfstring{$\to$}{to} v5 Measurement Scope}

\paragraph{Measurement Scope Note.}
The kernel selftest runs and micro/global overhead tables below
(Tables~\ref{tab:correctness}--\ref{tab:sanitizer-cmp}) were measured on the v4
prototype (\texttt{linux-next 7.3.0-rc5}, \texttt{next-20261001}, custom
clang/LLVM~24.0.0git \texttt{ca763b76e2fd}, rustc~1.101.0) under virtme-ng
(8~vCPUs, 1\,GB RAM, \texttt{CONFIG\_KASAN=y}), prior to switching to the
IR-only compiler contract, adding \texttt{pointee\_access\_c}, and adding
remote hardening. On the current v5 IR-only LLVM+Linux codebase, compiler
behavior and \texttt{-g} invariance are verified via LLVM regression tests
(\texttt{trace-args.ll}, \texttt{trace-ret.ll},
\texttt{trace-args-debug-invariance.ll}, and \texttt{trace-pointee-access.ll}),
Clang driver tests, \texttt{compiler-rt} \texttt{-O0}/\texttt{-O2 -g} tests, and
our kernel-wide IR pointer-snapshot audit across x86-64, AArch64, and Rust builds
(Section~\ref{sec:ir-audit}). Qualitatively, dropping compile-time DWARF field
tables shrinks \texttt{.rodata} growth while trading unannotated C entry
expansion for mid-call \texttt{0xA} access tracking
(\texttt{CONFIG\_KCOV\_DATAFLOW\_POINTEE\_ACCESS}). Re-running the full KVM
selftest and overhead matrix under v5 is scheduled as future empirical work.

In the baseline configuration, the kernel uses the selftest fragment
(\texttt{tools/\allowbreak testing/\allowbreak selftests/\allowbreak kcov\_dataflow/\allowbreak config}):
\begin{itemize}[nosep]
  \item \texttt{CONFIG\_KCOV=y}
  \item \texttt{CONFIG\_KCOV\_DATAFLOW\_ARGS=y}
  \item \texttt{CONFIG\_KCOV\_DATAFLOW\_RET=y}
  \item \texttt{CONFIG\_KCOV\_DATAFLOW\_INSTRUMENT\_ALL=y}
  \item \texttt{CONFIG\_KCOV\_DATAFLOW\_NO\_INLINE=y}
  \item \texttt{CONFIG\_DEBUG\_INFO\_DWARF5=y}, \texttt{CONFIG\_DEBUG\_FS=y}
  \item \texttt{CONFIG\_RUST=y}, \texttt{CONFIG\_MODULES=y}
  \item \texttt{CONFIG\_ANDROID\_BINDER\_IPC\_RUST=y} (binderfs comes with the
        driver; there is no longer a separate C binder or
        \texttt{CONFIG\_ANDROID\_BINDERFS} in this tree)
\end{itemize}
plus \texttt{CONFIG\_KASAN=y} for the sanitizer comparisons.

The v5 test suite in \texttt{tools/testing/selftests/kcov\_dataflow/} comprises
five kernel modules (3~C, 2~Rust) plus two user-space tests, driven by a KTAP
runner (note that \texttt{pointee\_access\_c} was added in v5 after the recorded
v4 run in Table~\ref{tab:correctness}):
\begin{itemize}[nosep]
  \item \texttt{eight\_struct\_args\_c} and
        \texttt{eight\_struct\_args\_rust}: 1--8 struct-pointer arguments,
        value-nested and pointer-linked struct towers on stack,
        \texttt{kmalloc} and \texttt{vmalloc} memory, pointer forwarding, and
        a 32-byte struct return lowered to indirect \texttt{sret}. Under v5's
        IR-only rule, plain C pointers and raw Rust \texttt{*const} pointers
        record only their addresses at entry, while \texttt{sret} captures all
        32 returned bytes.
  \item \texttt{pointee\_access\_c} (added in v5): exercises
        \texttt{CONFIG\_KCOV\_DATAFLOW\_POINTEE\_ACCESS}
        (\texttt{0xA} load/store records) through plain C pointers, including
        \texttt{container\_of()} negative offsets, spinlock-guarded loads,
        \texttt{READ\_ONCE}/\texttt{WRITE\_ONCE} (\texttt{\_VOLATILE}), and
        bit-field decoding via DWARF.
  \item \texttt{rust\_ffi\_contract}: an FFI contract violation---
        \texttt{ffi\_alloc\_buf(struct ffi\_alloc *alloc, ...)} returns 0 but
        leaves \texttt{alloc->buffer} \texttt{NULL}, and
        \texttt{ffi\_check\_result(struct ffi\_alloc alloc[static 1])} (IR
        \texttt{dereferenceable(24)}) captures the 24-byte pointee snapshot
        showing \texttt{buffer=NULL} and returns \texttt{-EFAULT}.
  \item \texttt{rust\_kworker\_remote}: \texttt{kcov\_df\_remote\_start()} /
        \texttt{stop()} from kworker context, capturing three phases of work
        queued on \texttt{system\_wq} through a published remote handle.
  \item \texttt{user\_ioctl} and \texttt{binderfs}: the ioctl interface and
        binder ioctls under active recording, both as TAP tests.
\end{itemize}
\texttt{test\_modules.py} loads each module, triggers it with recording
active, and checks the captured arguments, expanded struct fields, and return
values against the values the module uses; \texttt{trigger-view.py} is the
interactive viewer it is built on, with kallsyms/\texttt{addr2line}
symbolization and a \texttt{--remote} capture mode.
Source code is available at:
kernel\footnote{\url{https://github.com/yskzalloc/linux}},
LLVM\footnote{\url{https://github.com/yskzalloc/llvm-project}}, and
rustc\footnote{\url{https://github.com/yskzalloc/rust}}.

\subsection{Correctness Verification}

The suite is self-checking: \texttt{test\_modules.py} loads each module,
triggers it with recording active, and compares the captured records against
the values the module uses---25 assertions over argument indices, expanded
struct fields, object addresses, sizes, and return values.
Table~\ref{tab:correctness} reports a recorded run of the configuration in
Section~\ref{sec:eval} (kernel
\texttt{7.3.0-rc5-next-20261001+}, clang \texttt{24.0.0git}
\texttt{ca763b76e2fd}, LLD~24), x86-64 under KVM; Table~\ref{tab:correctness-tools}
places each scenario against the tools it would otherwise need.

\begin{table}[t]
\centering
\caption{Selftest results, all passing on the recorded run.}\label{tab:correctness}
\small
\begin{tabular}{@{}llc@{}}
\toprule
\textbf{Test} & \textbf{What is checked} & \textbf{Result} \\
\midrule
\texttt{user\_ioctl}      & 9 ioctl cases: sizing, double-init, mmap     & 9/9 \\
                          & ordering, enable/disable, records present    &     \\
\texttt{binderfs}         & binder ioctls under recording, 6185 words    & 3/3 \\
                          & captured, every record well-formed           &     \\
\texttt{rust\_ffi\_}      & 4 arg indices, \texttt{\{0,0,0,0\}} at entry,  & pass \\
\texttt{contract}         & \texttt{(256,16,1)}, ret 0, \texttt{\{NULL,0x110,0,0\}}, \texttt{-EFAULT} & \\
\texttt{eight\_struct\_}  & flat/value-nested/pointer-linked fields,      & pass \\
\texttt{args\_c}          & all return values, 32-byte \texttt{sret} return & \\
\texttt{eight\_struct\_}  & same expectations via \texttt{\#[repr(C)]}      & pass \\
\texttt{args\_rust}       & Rust twins                                     & \\
\texttt{rust\_kworker\_}  & records produced in kworker context arrive     & pass \\
\texttt{remote}           & in the enabling task's buffer                  & \\
\bottomrule
\end{tabular}
\end{table}

\begin{table}[t]
\centering
\caption{Which tool can observe each selftest scenario.}\label{tab:correctness-tools}
\small
\begin{tabular}{@{}lccc@{}}
\toprule
\textbf{Scenario} & \textbf{drgn} & \textbf{KASAN} & \textbf{\toolname{}} \\
\midrule
Struct fields of a C argument    & \cmark~(C) & N/A & \cmark \\
Struct fields of a Rust argument & \xmark & N/A & \cmark \\
Arguments 7 and 8                & \cmark~(C) & N/A & \cmark \\
Indirect (\texttt{sret}) return value & partial & N/A & \cmark \\
Postcondition violation, no crash & \xmark & \xmark & \cmark \\
Values seen inside a kworker     & \xmark & N/A & \cmark \\
\bottomrule
\end{tabular}
\end{table}

\noindent KASAN is marked N/A throughout because none of these scenarios is a
memory-safety violation---which is the point. \texttt{drgn} requires a crash
dump and reads C locals only where DWARF variable locations survive; it fails
outright on Rust frames at \texttt{-O2}.

\subsection{Information Gain Over Existing Tools}

Table~\ref{tab:infogain} compares the observability provided by \toolname{}
against KASAN reports and \texttt{drgn} vmcore analysis.

\begin{table}[t]
\centering
\caption{Information gain comparison.}\label{tab:infogain}
\small
\begin{tabular}{@{}lccc@{}}
\toprule
\textbf{Observable} & \textbf{KASAN} & \textbf{drgn} & \textbf{\toolname{}} \\
\midrule
Which argument corrupted & \xmark & \cmark~(C) & \cmark \\
Pre-corruption value & \xmark & partial & \cmark \\
Post-corruption value & \xmark & \cmark~(C) & \cmark \\
Cross-function propagation & \xmark & manual & \cmark \\
Rust module arguments & N/A & \xmark & \cmark \\
No-crash semantic bugs & \xmark & \xmark & \cmark \\
\bottomrule
\end{tabular}
\end{table}

KASAN reports the \emph{address} of a violation but not which argument carried
the bad value or which caller passed it. \texttt{drgn} can read C locals from a
vmcore but fails entirely on Rust frames compiled at \texttt{-O2} (DWARF
variable locations are elided). \toolname{} captures both C and Rust arguments
at runtime regardless of optimization level.

\subsection{Performance Overhead}

\paragraph{Measurement provenance.} Both measurements below were taken on the
toolchain of Section~\ref{sec:eval} by \texttt{bench.sh} in the repository,
which records the kernel, clang, rustc and commit of every run alongside the
numbers. An earlier version of this paper carried figures from linux-next
7.1.0-rc6 with clang/LLVM~23, measured against a micro-benchmark module that
was never carried into the in-tree suite, and predating the ABI of
Section~\ref{sec:llvm-pass}---so they could neither be reproduced nor
compared. Where the two disagree it is noted, because the disagreements are
informative.

Absolute values are not comparable with that earlier table: the kernel
version, the configuration and the syscall mix all differ. The percentages
are.

\paragraph{Cost per recorded value.} What has been re-measured on the current
toolchain is the quantity that transfers between workloads: the cost of
storing one record. Driving \texttt{eight\_struct\_args\_c}'s trigger on
kernel \texttt{7.3.0-rc5-next-20261001+} with clang~\texttt{24.0.0git}
(\texttt{ca763b76e2fd}) and rustc~1.101.0, x86-64 under KVM, with only that
module carrying arg/ret instrumentation, gives
$\mathbf{18}$--$\mathbf{21}$\,\textbf{ns per stored record} across four runs
(17.1, 18.3, 20.6 and 20.9\,ns; the overhead on the same runs ranged 65--75\%).
The spread is the honest precision available from a guest with eight vCPUs
sharing a host, and it is reported rather than narrowed to the most favourable
pair. The benchmark derives this by counting the records a run
actually stored rather than by assuming how many calls the trigger makes---an
assumption that was wrong by a factor of 16 when checked, which would have
inflated the per-record figure by the same factor.

The percentage overhead is deliberately not promoted to a headline, because it
is a property of record \emph{density} rather than of the instrumentation: the
same $\sim$20\,ns costs +75\% on this trigger, which crosses 10{,}587 recorded
boundaries and comparisons per iteration, and far less on a workload that
crosses fewer. Table~\ref{tab:perf} therefore reports the density beside the
latency. An earlier version of this paper quoted +8.3\% at 44 callbacks per
iteration with no density stated, which made the figure impossible to compare
against anything---including its own successor.

\begin{table}[t]
\centering
\caption{Cost of recording, measured with \texttt{bench.sh micro} on kernel
\texttt{7.3.0-rc5-next-20261001+}, clang~\texttt{24.0.0git}, x86-64 under KVM.
Only \texttt{eight\_struct\_args\_c} carries arg/ret instrumentation; the
comparison records come from \texttt{CONFIG\_KCOV\_ENABLE\_COMPARISONS}
sharing the same buffer.}\label{tab:perf}
\small
\begin{tabular}{@{}lrr@{}}
\toprule
\textbf{Configuration} & \textbf{Latency} & \textbf{Overhead} \\
\midrule
Instrumented, not recording & 291.3\,$\mu$s/iter & --- \\
\toolname{} recording ON    & 509.8\,$\mu$s/iter & +75.0\% \\
\midrule
\multicolumn{3}{@{}l@{}}{\emph{density:} 10{,}587 records/iter --- 1{,}199 argument
and return, rest comparison} \\
\multicolumn{3}{@{}l@{}}{\emph{derived:} 20.6\,ns per stored record on this run;
18--21\,ns across four} \\
\bottomrule
\end{tabular}
\end{table}

The per-record cost is dominated by the slot reservation---a
\texttt{READ\_ONCE}/\texttt{WRITE\_ONCE} validate-then-commit of the word count
at \texttt{area[0]}, with no atomic and no lock on the local path (portable to
ARM64)---plus one \texttt{copy\_from\_kernel\_nofault()} per struct field. A
record that carries its value in a register performs no memory read at all.

Critically, non-instrumented code paths incur \emph{zero} overhead---the
compiler emits no callbacks for modules without the
\texttt{KCOV\_DATAFLOW} flag.

\paragraph{Global Enablement Overhead.}
With \texttt{CONFIG\_KCOV\_DATAFLOW\_INSTRUMENT\_ALL=y} and
\texttt{CONFIG\_KCOV\_DATAFLOW\_NO\_INLINE=n} (inlining preserved), the kernel
boots normally under virtme-ng. Table~\ref{tab:global-perf} quantifies the
overhead when all kernel objects are instrumented but recording is
\emph{not} enabled (the common idle state). These rows come from two full
kernel builds of the tree in Section~\ref{sec:eval}, differing only in
\texttt{INSTRUMENT\_ALL}, measured by \texttt{bench.sh global}.

\begin{table}[htbp]
\centering
\caption{Global instrumentation overhead, recording disabled. Two builds of
kernel \texttt{7.3.0-rc5-next-20261001+} with clang~\texttt{24.0.0git},
x86-64 under KVM.}\label{tab:global-perf}
\small
\begin{tabular}{@{}lrrr@{}}
\toprule
\textbf{Metric} & \textbf{Baseline} & \textbf{INSTRUMENT\_ALL} & \textbf{Overhead} \\
\midrule
vmlinux .text        & 54.9\,MB & 78.6\,MB & +43\% \\
vmlinux .rodata      & 12.3\,MB & 19.5\,MB & +58\% \\
vmlinux .data        & 9.9\,MB  & 9.9\,MB  & +0\% \\
bzImage              & 34.7\,MB & 44.0\,MB & +27\% \\
Time to userspace    & 11.0\,s  & 17.9\,s  & +62\% \\
Syscall latency      & 132.6\,$\mu$s/iter & 220.0\,$\mu$s/iter & +66\% \\
\bottomrule
\end{tabular}
\end{table}

The growth is in read-only and executable memory, not writable data.
\texttt{.rodata} takes +58\% (+7.2\,MB) because that is where the
\texttt{\_\_sancov\_offsets\_} field tables live: they are constant globals,
one per traced composite type. \texttt{.data} is unchanged to the byte, which
is the useful negative result---an earlier version of this paper attributed a
+44\% \texttt{.data} increase to those tables, which was wrong in both the
number and the mechanism. \texttt{.text} takes +43\% (+23.7\,MB) from the
per-parameter call setup planted at every frame start, and objtool's
\texttt{.call\_sites} grows +3.8\,MB alongside it, being a table indexed by
exactly those new calls.

The syscall figure (+66\%) is the cost of instrumentation alone: recording is
disabled, so each callback is a call and an early return. For
comparison, basic KCOV (\texttt{trace-pc}) adds only 5--10\% with
\texttt{INSTRUMENT\_ALL} since it emits a single call per edge with no argument
setup.

Global enablement is suitable for targeted investigation sessions and
short-lived fuzzing campaigns. For continuous fuzzing (e.g., multi-day
syzkaller runs), per-module instrumentation remains recommended.

\paragraph{Comparison with Kernel Sanitizers.}
Table~\ref{tab:sanitizer-cmp} contextualizes \toolname{}'s overhead against
established kernel sanitizers. All measurements include
\texttt{CONFIG\_KASAN=y} in the baseline (the standard fuzzing configuration).

\begin{table}[htbp]
\centering
\caption{Overhead comparison with kernel sanitizers. The \toolname{} row is measured (Table~\ref{tab:global-perf}); sanitizer ranges are from the literature.}\label{tab:sanitizer-cmp}
\small
\begin{tabular}{@{}lrl@{}}
\toprule
\textbf{Tool} & \textbf{Syscall Overhead} & \textbf{Instruments} \\
\midrule
Basic KCOV (trace-pc)       & +5--10\%   & 1 call/edge \\
KCOV + comparisons          & +15--30\%  & edges + cmp operands \\
\toolname{} (INSTRUMENT\_ALL) & +66\%    & per-arg + per-ret callbacks \\
KASAN (generic)             & +100--200\% & every memory access \\
KCSAN                       & +100--200\% & every memory access \\
KMSAN                       & +200--400\% & every uninitialized use \\
\bottomrule
\end{tabular}
\end{table}

\noindent The overhead of \toolname{} sits below a single
sanitizer (KASAN/KCSAN) and well below KMSAN. Since fuzzing kernels
routinely stack multiple sanitizers (KASAN + KCOV + UBSAN), adding dataflow
extraction fits naturally within the accepted performance budget.
The ``KCOV + comparisons'' row is not additive with ours: \toolname{} reuses
that same instrumentation rather than adding a second copy, so enabling
\texttt{CONFIG\_KCOV\_ENABLE\_COMPARISONS} alongside the dataflow modes costs
the comparison callbacks once and routes each operand pair to both collectors
(Section~\ref{sec:ringbuf}).
Notably, both builds in Table~\ref{tab:global-perf} have
\texttt{CONFIG\_KASAN=y}, so the +66\% is the \emph{additional} cost on top of
an already-sanitized kernel rather than a measurement against a bare one.

\subsection{v5 IR-Only Pointee Visibility Audit}\label{sec:ir-audit}

To measure precisely how much pointee data the strict IR-only snapshot contract
(\texttt{noundef dereferenceable(N)} / \texttt{dereferenceable\_or\_null(N)})
captures across real kernel code without guessing from debug info, we audited
Linux kernel builds compiled with the v5 LLVM pass
(Table~\ref{tab:ir-audit}):
\begin{itemize}[nosep]
  \item \textbf{x86-64 C pointer arguments:} 0 of 23{,}810 pointer arguments
        carry an IR dereferenceability contract and thus record pointee bytes
        at entry; every ordinary C \texttt{T *} pointer is recorded as its
        address only (unless passed \texttt{byval}/\texttt{sret} or declared
        \texttt{T p[static N]}).
  \item \textbf{AArch64 C pointer arguments:} 21 of 22{,}606 pointer arguments
        record entry pointee bytes, corresponding to structs over 16~bytes that
        the AArch64 ABI passes by reference with Clang-emitted
        \texttt{dereferenceable} attributes.
  \item \textbf{Kernel Rust object pointer arguments:} 318 of 756 (42\%)
        pointer arguments record entry pointee snapshots automatically, because
        \texttt{rustc} emits \texttt{noundef dereferenceable(size\_of::<T>())}
        on safe \texttt{\&T} and \texttt{\&mut T} references (and
        \texttt{dereferenceable\_or\_null} on \texttt{Option<\&T>}). Raw pointers,
        slices, and \texttt{\&} to \texttt{!Freeze} types (\texttt{Opaque<T>},
        \texttt{Cell}) remain address-only.
\end{itemize}
\begin{table}[t]
\centering
\caption{v5 IR-only entry pointee-snapshot coverage across Linux kernel builds
(fraction of pointer arguments with $\geq 1$ recorded pointee byte).}\label{tab:ir-audit}
\small
\begin{tabular}{@{}lrr@{}}
\toprule
\textbf{Kernel Build Scope} & \textbf{Snapshotted / Total Ptrs} & \textbf{Share} \\
\midrule
Linux x86-64 C pointer args   & 0 / 23{,}810  & 0.0\% \\
Linux AArch64 C pointer args  & 21 / 22{,}606 & 0.09\% \\
Linux Rust object pointer args & 318 / 756    & 42.1\% \\
\bottomrule
\end{tabular}
\end{table}

\noindent This measurement directly motivates the companion
\texttt{KCOV\_DF\_TYPE\_ACCESS} (\texttt{0xA}) extension
(Section~\ref{sec:deref}): while pure IR snapshots cover 42\% of Rust object
boundaries with zero manual annotations, tracing constant-offset loads and
stores performed through pointer arguments closes the 99.9\%--100\% plain-C
pointer visibility gap without introducing unsafe or unverified speculative reads.

\subsection{Triage Capability}

For each captured boundary, the ENTRY/RET record provides: (1)~which argument
carried a value, identified by \texttt{arg\_idx} and, within a struct, by
position in the field table; (2)~whether a struct field changed across a call,
visible as a difference between the ENTRY record of a callee and the ENTRY
record of the next function to receive the same object---this is what makes the
\texttt{rust\_ffi\_contract} postcondition violation legible; and (3)~the order
in which boundaries were crossed, from the 24-bit per-task sequence number, with
the PC identifying the call site. Resolving that PC with
\texttt{sancov -symbolize-dataflow} or \texttt{addr2line -i} maps the call back
to source-level parameter/member names and, for mid-function \texttt{0xA}
accesses inside an inlined helper, names the inlined site.

What the records do \emph{not} give is a call \emph{tree}: there is no caller
field and no depth counter, so \texttt{trigger-view.py} reconstructs nesting
heuristically from sequence order and symbol boundaries. A record also carries
no piece index, so two register pieces of one struct return are
indistinguishable from two calls to the same function
(Section~\ref{sec:eval}).

\subsection{Eight Struct Arguments, C and Rust}

To verify that \toolname{} captures Rust kernel module function arguments
natively (without the post-compilation pipeline), I built \texttt{rustc}
against the custom LLVM~24 containing the trace-args/ret passes. The kernel
build system passes \texttt{-Cllvm-args=-sanitizer-coverage-trace-args}
and \texttt{-Cllvm-args=-sanitizer-coverage-trace-ret} to rustc for modules
with \texttt{KCOV\_DATAFLOW\_module.o := y}.

The \texttt{eight\_struct\_args\_c} module and its Rust twin
\texttt{eight\_struct\_args\_rust} exercise three struct families, each with
eight members:
\begin{itemize}[nosep]
  \item \textbf{Flat} \texttt{s1..s8}: \texttt{sN} holds $N$ \texttt{u64}
        members side by side, and \texttt{sf\_N} takes $N$ struct pointers.
        This covers plain field expansion and, at \texttt{sf\_7}/\texttt{sf\_8},
        arguments the x86-64 ABI passes on the stack rather than in the first
        six registers.
  \item \textbf{Value-nested} \texttt{st1..st8}: \texttt{stN} embeds every
        smaller struct \emph{by value}, so \texttt{st8} is eight levels deep.
        Exercised on the stack.
  \item \textbf{Pointer-linked} \texttt{stp1..stp8}: every member is a pointer
        to a separately allocated object, so the nesting runs through the heap.
        Exercised with both \texttt{kmalloc} and \texttt{vmalloc} memory.
\end{itemize}
All fields carry the values $0\mathtt{x}11, 0\mathtt{x}22, \ldots$ (Figure~\ref{fig:eight-args}), and
\texttt{test\_modules.py} asserts each expanded field and every return value
(\texttt{sf\_N} returns $\sum_{k=1}^{N} 0\mathtt{x}11k$; the nested families
return \texttt{field0} plus the recursive sums of the embedded structs).

The two nested families are where the shallow-capture boundary of
Section~\ref{sec:deref} becomes observable rather than theoretical: the
value-nested tower yields one \emph{(offset, size)} pair per top-level member,
with an embedded struct clamped to its first word, and the pointer-linked tower
yields the pointer values rather than the objects behind them. The tests assert
exactly that, so the limitation is pinned down by the suite instead of being
left as an assumption.

\paragraph{Indirect struct returns.} \texttt{sf\_ret\_struct()} returns
\texttt{struct s4} by value. At $4 \times$~\texttt{u64}~=~32 bytes that is over
the register-return limit on both x86-64 and arm64, so the ABI returns it
indirectly through a caller-provided buffer---and \texttt{trace-ret} reports
that buffer the way it reports a struct argument: one value word per field,
word~[2] of the record holding the address the fields were read from, and
\texttt{size} the whole object rather than one field:
\begin{lstlisting}[language=,numbers=none,columns=fullflexible,basicstyle=\ttfamily\scriptsize,frame=none]
[RET] seq=.. sf_ret_struct+0x91 ret(32) @0xffff888000003000 = {0x11,0x11,0x22,0x33}
\end{lstlisting}
The runner asserts all four fields, the object size, \emph{and} that word~[2]
is non-zero---a scalar return leaves it zero, so that word is what
distinguishes the two record shapes. Checking only the first value word would
pass equally for a scalar return. \texttt{rsf\_ret\_struct()} takes the same
path from an \texttt{extern "C"} Rust function with a \texttt{\#[repr(C)]}
return type, and both modules are checked through one shared expectation---the
point of having both. A struct small enough to come back in registers
($\leq$~16 bytes here) has no address to report from and instead arrives as one
record per register piece with no field table; because the piece records carry
no piece index, a consumer cannot tell two pieces of one return from two calls
to the same function. That path is covered by the LLVM IR tests rather than by
the kernel suite.

\paragraph{\texttt{\#[repr(C)]} is load-bearing.} It is what makes the field
offsets the compiler reports match the C layout. A default-repr Rust struct may
be reordered, and the reported offsets would then describe that layout
instead---still correct, but not comparable with the C module field by field.

\subsection{Remote Capture from Kworker Context}

The \texttt{rust\_kworker\_remote} module exercises the remote path: the
trigger queues a work item on \texttt{system\_wq} whose three phases
(populate, update, drain of a composite store of red-black trees) run inside
\texttt{kcov\_df\_remote\_start()}/\texttt{stop()} on handle~1, which the
runner publishes with \texttt{KCOV\_DF\_REMOTE\_ENABLE}. The records land in
the enabling task's buffer even though they were produced by a kernel thread
that never entered a syscall, which is the gap that per-task-only collection
leaves open and that \texttt{ftrace}'s global patching fills only by also
capturing every other task on the system.

\begin{figure}[t]
\centering
\begin{tikzpicture}[
  node distance=0.12cm,
  entry/.style={font=\ttfamily\scriptsize, anchor=west, text=blue!70!black},
  ret/.style={font=\ttfamily\scriptsize, anchor=west, text=red!60!black},
  note/.style={font=\tiny\itshape, anchor=west, text=black!60},
]
\node[entry] (e1) at (0,0)    {sf\_2(arg0=\colorbox{orange!20}{\{0x11\}}, arg1=\colorbox{orange!20}{\{0x11, 0x22\}})};
\node[ret]   (r1) at (0.3,-0.35) {\colorbox{green!15}{0x33} = sf\_2()};
\node[entry] (e2) at (0,-0.8) {sf\_8(arg0..arg5 in registers, \colorbox{yellow!20}{arg6}, \colorbox{yellow!20}{arg7} on the stack)};
\node[note]  (n2) at (0.3,-1.12) {all eight reported; the ABI's register/stack split is invisible to the pass};
\node[entry] (e3) at (0,-1.6) {stf\_3(arg0=\colorbox{orange!20}{\{0x33, 0x11, 0x11\}})};
\node[note]  (n3) at (0.3,-1.92) {value-nested: one pair per top-level member, embedded struct clamped to its first word};
\node[entry] (e4) at (0,-2.4) {stpf\_3(arg0=\colorbox{orange!20}{\{0xffff888.., 0xffff888.., 0xffff888..\}})};
\node[note]  (n4) at (0.3,-2.72) {pointer-linked: the pointers, not the heap objects behind them};
\node[ret]   (r5) at (0,-3.2) {\colorbox{green!15}{\{0x11, 0x11, 0x22, 0x33\}} = sf\_ret\_struct() \quad ret(32) @0xffff888000003000};
\node[note]  (n5) at (0.3,-3.52) {indirect \texttt{sret} return: word [2] non-zero, size is the whole 32-byte object};

\node[font=\tiny, anchor=west] at (0,-4.0) {%
  \colorbox{yellow!20}{scalar args}\quad
  \colorbox{orange!20}{expanded struct fields}\quad
  \colorbox{green!15}{return values}};
\end{tikzpicture}
\caption{Representative capture from the v4 prototype on
\texttt{eight\_struct\_args\_c}; the Rust twin (\texttt{rsf\_*},
\texttt{rstf\_*}, \texttt{rstpf\_*}) checked the same expectations. Under v5's
IR-only contract, plain C pointers and raw Rust pointers record pointer values
at entry (and constant-offset accesses under \texttt{POINTEE\_ACCESS}), while
indirect \texttt{sret} returns (\texttt{sf\_ret\_struct}) still capture all
32~returned bytes.}\label{fig:eight-args}
\end{figure}

In the v4 baseline run, all 8 argument values and C/Rust struct fields were
captured with matching \texttt{sret} expectations; under v5's strict IR-only
rule, \texttt{eight\_struct\_args\_c} and \texttt{eight\_struct\_args\_rust}
pin down that plain C pointers (\texttt{sf\_N}) and Rust raw pointers
(\texttt{*const SN}) emit addresses at entry rather than unverified pointee
reads, while \texttt{pointee\_access\_c} captures plain-C member loads/stores
via \texttt{0xA} records.

\subsection{FFI Contract Auditing}

The \texttt{rust\_ffi\_contract} module demonstrates \toolname{}'s primary
auditing use case: a \emph{postcondition} violation at a boundary where the
caller has no way to detect it. The callee reports success and leaves the
caller's struct in a state its contract forbids; the next boundary receives that
struct with an explicit IR contract (\texttt{alloc[static 1]} $\to$
\texttt{dereferenceable(24)}), surfacing the broken postcondition in its
24-byte entry snapshot (Figure~\ref{fig:ffi-contract}).

\begin{figure}[t]
\centering
\begin{tikzpicture}[
  node distance=0.4cm,
  funcbox/.style={rectangle, draw, thick, rounded corners, fill=blue!8, minimum width=7.5cm, minimum height=0.6cm, font=\scriptsize\bfseries, align=left, inner sep=4pt},
  innerbox/.style={rectangle, draw, rounded corners, fill=purple!8, minimum width=7.5cm, minimum height=0.6cm, font=\scriptsize\bfseries, align=left, inner sep=4pt},
  retbox/.style={rectangle, draw, rounded corners, fill=green!10, minimum width=7.5cm, minimum height=0.45cm, font=\scriptsize, align=left, inner sep=3pt},
  instlabel/.style={rectangle, draw=green!60!black, fill=green!10, font=\tiny\bfseries, inner sep=2pt, rounded corners=1pt},
  arrow/.style={-{Stealth[length=2mm]}, thick},
]
\node[funcbox] (rcall) {[ENTRY] ffi\_alloc\_buf(\colorbox{yellow!25}{alloc=0xffff..}, \colorbox{yellow!25}{256}, \colorbox{yellow!25}{16}, \colorbox{yellow!25}{1})};
\node[instlabel, anchor=south east] at ([xshift=-2pt, yshift=2pt]rcall.north east) {4 args (plain ptr)};

\node[retbox, below=0.4cm of rcall] (cret) {[RET] \quad ffi\_alloc\_buf() = \colorbox{green!20}{0} \quad -- claims success};

\node[innerbox, below=0.72cm of cret] (ccall) {[ENTRY] ffi\_check\_result(\colorbox{orange!25}{[static 1] $\to$ \{.buffer=NULL, .data\_size=0x110, 0, 0\}})};
\node[instlabel, anchor=south east] at ([xshift=-2pt, yshift=2pt]ccall.north east) {deref(24) violation};

\node[retbox, below=0.4cm of ccall, fill=red!12] (rret) {[RET] \quad ffi\_check\_result() = \colorbox{red!25}{-EFAULT}};

\draw[thick, red!60, dashed] ([xshift=-0.3cm, yshift=0.46cm]ccall.north west) -- ([xshift=0.3cm, yshift=0.46cm]ccall.north east)
  node[pos=1, right, font=\tiny\bfseries, text=red!60] {next boundary};
\end{tikzpicture}
\caption{FFI contract auditing in \texttt{rust\_ffi\_contract}.
\texttt{ffi\_alloc\_buf()} receives a plain C pointer (address logged at entry)
and returns~0, claiming success. The checker receives the same object declared
as \texttt{alloc[static 1]} (\texttt{dereferenceable(24)}), so its 24-byte
entry snapshot (symbolized offline via DWARF) exposes
\texttt{buffer}~=~\texttt{NULL} before \texttt{-EFAULT} is
returned.}\label{fig:ffi-contract}
\end{figure}

The \texttt{rust\_ffi\_contract} module encodes this pattern directly:
\texttt{ffi\_alloc\_buf(struct ffi\_alloc *alloc, ...)}'s documented contract is
``returns 0 $\implies$ \texttt{alloc->buffer} is valid,'' yet its async path with
\texttt{free\_async == 0} returns 0 while leaving \texttt{buffer} \texttt{NULL}.
Because \texttt{alloc} is a plain C pointer, \texttt{ffi\_alloc\_buf} records
its pointer address and scalars \texttt{(256, 16, 1)} plus return \texttt{0}.
Next, \texttt{ffi\_check\_result(struct ffi\_alloc alloc[static 1])} declares a
static-size parameter, producing IR \texttt{dereferenceable(24)}: \toolname{}
records all 24 pointee bytes (decoded offline via DWARF to \texttt{\{NULL,
0x110, 0, 0\}}) and the \texttt{-EFAULT} return. No crash occurs, no KASAN
report fires, and edge coverage is identical to the correct path.

A verifier needs both boundaries to see it: the return value alone says
success, and the struct alone says nothing about what was promised. This is the
smallest complete instance of the postcondition checking formalized in
Section~\ref{sec:contracts}.

Neither \texttt{drgn} (DWARF variable locations elided at \texttt{-O2}) nor
\texttt{ftrace}'s function-args printing (which would need the
\texttt{struct ffi\_alloc} fields, not the pointer) reports this without
writing a target-specific kprobe or eBPF handler per function.

\section{Case Study: Runtime Boundary Auditing}\label{sec:realworld}

I applied \toolname{} to production kernel code to evaluate whether
boundary data-flow records provide actionable information beyond what
static review, KASAN, or edge coverage alone can surface.
I emphasize that the goal is \emph{demonstrating the methodology},
not claiming novel vulnerability discovery: several observations below
were subsequently clarified by Rust kernel community members as intentional behavior.

\subsection{Methodology}

For each target subsystem, I follow a four-step process:
\begin{enumerate}[nosep]
  \item Enable \texttt{KCOV\_DATAFLOW} on the target module
        (\texttt{KCOV\_DATAFLOW\_module.o := y}).
  \item Write a user-space program exercising boundary inputs (edge values,
        NULL pointers, maximum sizes).
  \item Capture the dataflow records via \texttt{mmap()} during execution.
  \item Run an \emph{automated contract verifier} that checks each captured
        $(\mathit{args}, \mathit{ret})$ pair against a contract database
        (Section~\ref{sec:contracts}).
\end{enumerate}

The key insight is that steps~3--4 require no human judgment for known
contracts: the verifier mechanically flags violations by comparing captured
values against pre/postcondition predicates. Only \emph{novel} violations
(functions without existing contracts) require manual analysis and maintainer
discussion. This shifts the workflow from ``report every observation and
wait for feedback'' to ``automatically verify known contracts; escalate only
confirmed anomalies.''

\paragraph{Why compile-time checking is insufficient.}
Rust's type system and C's \texttt{\_\_attribute\_\_} annotations enforce
structural constraints (correct types, lifetime bounds) but cannot verify
\emph{value} constraints that depend on runtime kernel state. A function
signature \texttt{alloc\_buf(size: u32)} type-checks for any \texttt{u32},
but at runtime the valid range depends on non-deterministic conditions:
available slab memory, per-task resource limits, concurrent allocations, and
RCU grace period state. These conditions are unknowable at compile time.
Boundary contract verification fills this gap: by capturing the \emph{actual}
argument values observed at runtime and checking them against value predicates
($\mathit{size} \leq \mathit{pool\_capacity}$), we verify invariants that
no static analysis or type system can express.

\emph{Critical distinction:} \toolname{} captures \emph{what the kernel
actually does} at runtime. Whether that behavior constitutes a bug requires
human judgment and maintainer confirmation---the tool provides evidence,
not verdicts.

\paragraph{Violation detection criteria.} A captured record is flagged as a
potential violation when any of the following hold:
\begin{enumerate}[nosep]
  \item An ENTRY argument value falls outside the documented or inferred
        valid range for that parameter (e.g., size $>$ buffer capacity).
  \item A RET value differs from what the precondition guarantees (e.g.,
        \texttt{from\_errno(-4096)} returns \texttt{-22} instead of
        propagating the input).
  \item A struct field changes between ENTRY and RET in a function documented
        as read-only with respect to that field.
  \item A pointer argument is \texttt{NULL} or in the \texttt{ERR\_PTR} range
        at a call site whose callee performs unconditional dereference.
\end{enumerate}
These criteria are applied offline against the captured buffer; the kernel
itself makes no policy decisions during recording.

\subsection{Observations and Kernel Community Responses}

\paragraph{Binder \texttt{SET\_MAX\_THREADS} (reported, not a bug).}
\toolname{} captured \texttt{BINDER\_SET\_MAX\_THREADS(0xffffffff)} being
accepted without bounds validation at
\texttt{drivers/android/binder/process.rs}'s ioctl dispatch. I initially
hypothesized this could
enable OOM via unbounded thread creation. Kernel maintainer Alice Ryhl
confirmed that this value does \emph{not} bypass \texttt{RLIMIT\_NPROC}:
the kernel enforces process-level thread limits independently. The recorded
dataflow correctly showed the ioctl accepting the value; the \emph{interpretation}
that this was exploitable was incorrect.

\paragraph{Binder \texttt{BC\_ENTER\_LOOPER} duplicate (reported, not a bug).}
Dataflow records showed a duplicate \texttt{BC\_ENTER\_LOOPER} command being
accepted at \texttt{drivers/android/binder/thread.rs}'s command loop. Rust
Kernel Community review confirmed the operation is bit-level idempotent
(\texttt{looper |= BINDER\_LOOPER\_STATE\_ENTERED}) and produces no corruption.

\noindent Both observations are on the \emph{Rust} binder: this tree carries no
C binder driver, only
\texttt{drivers/android/binder/}, gated on
\texttt{CONFIG\_ANDROID\_BINDER\_IPC\_RUST}. That is the substantive part
even though the findings themselves were refuted---the values were read out of
Rust frames at \texttt{-O2}, which is where \texttt{drgn} stops and where
ftrace's BTF-based printing cannot decode the arguments.

\paragraph{\texttt{Error::from\_errno} boundary (confirmed, pre-existing TODO).}
The dataflow record captured \texttt{from\_errno(-4096)} returning \texttt{-22}
(EINVAL). The source already contains a \texttt{// TODO: Make it a WARN\_ONCE}
comment acknowledging this edge case. This demonstrates \toolname{}'s ability
to surface known-but-unresolved issues through runtime evidence rather than
source inspection.

\subsection{Rust Kernel Core API Verification}

I exercised \texttt{rust/kernel/} APIs with boundary inputs to verify
correct behavior (not to find bugs). The dataflow records confirmed:

\begin{itemize}[nosep]
  \item \textbf{KVec:} \texttt{reserve(usize::MAX)} returns
        \texttt{AllocError}; \texttt{push\_within\_capacity}
        on a full vec returns the rejected value.
  \item \textbf{RBTree:} Duplicate key insert replaces the existing value
        (documented behavior, differs from \texttt{std::HashMap::try\_insert}).
  \item \textbf{Arc:} \texttt{into\_unique\_or\_drop} with multiple refs
        correctly drops; refcount operations captured at \texttt{-O2}
        where \texttt{drgn} cannot observe them.
  \item \textbf{Page:} Cross-page boundary reads rejected
        (offset+len $>$ 4096 $\to$ EINVAL).
  \item \textbf{UserSlice:} NULL, kernel-range, and unmapped addresses all
        return EFAULT correctly.
\end{itemize}

\noindent These results demonstrate that \toolname{} is equally valuable for
\emph{building confidence in correct implementations} as for identifying
potential issues. The dataflow records provide reproducible evidence that
APIs handle boundary inputs as intended---evidence that static review alone
cannot provide at \texttt{-O2}.

\paragraph{Reproducibility caveat.} These observations were made with ad-hoc
modules that were not carried into the in-tree selftest suite of
Section~\ref{sec:eval}. They are therefore not reproducible from the published
tree as it stands; only the four suite modules are. The one exception is
\texttt{RBTree}, which \texttt{rust\_kworker\_remote} exercises, though for the
remote-capture path rather than for duplicate-key semantics. Carrying the
remaining API probes into the suite is outstanding work.

\subsection{Correction Note}

In version~1 of this paper, I identified several potential vulnerabilities
in the Rust Android Binder driver and Rust kernel core APIs, assigning severity
labels (High, Critical) prior to maintainer verification. Following review
and discussion with the Linux kernel maintainers, these specific findings
were determined to be not exploitable under standard configurations or
addressed by existing kernel mitigations:

\begin{itemize}[nosep]
  \item \texttt{BINDER\_SET\_MAX\_THREADS(0xffffffff)}: not exploitable---
        \texttt{RLIMIT\_NPROC} enforcement is independent and prevents
        unbounded thread creation regardless of the stored value.
  \item \texttt{BC\_ENTER\_LOOPER} duplicate: not a corruption vector---
        the operation is bit-level idempotent.
  \item Rust API behavioral observations (e.g., \texttt{RBTree} duplicate
        key semantics): documented intentional behavior, not bugs.
\end{itemize}

\noindent Consequently, these findings have been removed from this version.
The system design, methodology, and the runtime boundary auditing framework
presented in the preceding sections remain valid. Verified findings will be
incorporated step by step in subsequent versions as maintainer review
completes.

\section{Why ftrace Does Not Replace Boundary Data-Flow}\label{sec:ftrace}

A natural question is whether existing tracing infrastructure (\texttt{ftrace},
\texttt{kprobes}, eBPF) could achieve similar results. It can reach part of the
way, and the gaps are architectural rather than incidental.

\paragraph{Global text patching, not per-task instrumentation.}
\texttt{ftrace}, \texttt{kprobes}, and eBPF all patch global kernel text, so a
probe fires on every task on every CPU. Per-task selection is a post-trigger
filter applied to a shared stream, not a property of the instrumentation.
\toolname{} records into the enabling task's own buffer, and follows that task
through \texttt{execve}---which is what makes a captured record a reproducible
per-input artifact rather than a slice of system-wide activity.

\paragraph{Six arguments, and no more.} ftrace's function-args feature is
capped at \texttt{FTRACE\_REGS\_MAX\_ARGS}, which is 6. Scanning the
\texttt{next-20260831} tree, 8{,}195 of 807{,}481 distinct functions (1.01\%)
declare more than six parameters, and ftrace truncates the 7th onward for every
one of them; the tail is long, reaching 33 parameters. Restricting the count to
functions that actually survive as a text symbol in one representative
\texttt{vmlinux}---the real ftrace targets---still leaves 668. The dataflow
callbacks have no such cap: they report one call per declared parameter.

\paragraph{Rust is only partially reachable.} \texttt{rustc} does emit
\texttt{\_\_fentry\_\_} stubs---\texttt{-Zinstrument-mcount} accepts
\texttt{fentry}, \texttt{fentry-record}, and \texttt{fentry-nop-record} on
supported targets---so the claim that Rust objects are simply unpatchable is
wrong. The gap is one level up: idiomatic Rust types (generics, slices, trait
objects) do not lower to the \texttt{repr(C)} shapes that
\texttt{pahole}/BTF-based argument printing expects, so even where the probe
site exists, the argument values are not decodable by the BTF machinery.
\toolname{} captures arguments directly from LLVM IR---preserving values at
\texttt{-O2} and snapshotting Rust \texttt{\&T}/\texttt{\&mut T} pointees via
their IR \texttt{dereferenceable} attributes---and relies on DWARF only for
optional offline symbolization.

\paragraph{Inlining Trade-Offs.} A callee the inliner merged
away leaves no \texttt{call} for ftrace to patch, and no
\texttt{\_\_fentry\_\_} stub; reaching it requires an explicit
\texttt{noinline} or \texttt{\#[inline(never)]} in the source. Under v5's
strict IR-only design, \toolname{} likewise instruments post-inlining IR
functions so \texttt{-g} cannot alter codegen; when an investigation requires
observing internal helper boundaries,
\texttt{CONFIG\_KCOV\_DATAFLOW\_NO\_INLINE} enables \texttt{-fno-inline} per
opted-in module without runtime probing edits.

\paragraph{Rust symbol mangling.} Rust function names are mangled into
unreadable strings (e.g., \texttt{\_RNvMs3\_NtCsgHKxaxYGgF6\_11rust\_binder...}).
Setting ftrace filters or kprobe targets requires these exact names, whereas
the dataflow pass instruments by compiler flag and needs no symbol list.

\paragraph{\toolname{} operates at LLVM IR level.} My pass runs on LLVM IR
\emph{before} target codegen---the common representation shared by both
\texttt{clang} (C) and \texttt{rustc} (Rust). At this level, function arguments
are explicit SSA values and attributes, and inserted \texttt{notail},
\texttt{nomerge} callbacks survive optimization without forcing stack
allocations. Table~\ref{tab:ftrace} summarizes the comparison.

\begin{table}[t]
\centering
\caption{Comparison: ftrace/kprobes vs.\ \toolname{}.}\label{tab:ftrace}
\small
\begin{tabular}{@{}lcc@{}}
\toprule
\textbf{Capability} & \textbf{ftrace} & \textbf{\toolname{}} \\
\midrule
Probe site in Rust objects & \cmark & \cmark \\
Decode idiomatic Rust arguments & \xmark & \cmark \\
Arguments beyond the 6th & \xmark & \cmark \\
IR object / pointee / \texttt{0xA} capture & \xmark & \cmark \\
No symbol-mangling filter needed & \xmark & \cmark \\
Per-task isolation by construction & \xmark & \cmark \\
Zero source modification & \cmark & \cmark \\
Works at \texttt{-O2} & \cmark & \cmark \\
No recompilation needed & \cmark & \xmark \\
\bottomrule
\end{tabular}
\end{table}

\noindent The last row is the honest cost of the approach: ftrace works on a
shipped kernel, while \toolname{} requires a build with the instrumentation
flags.

\section{Boundary Contract Runtime Verification}\label{sec:contracts}

I observe that \toolname{} enables a new form of \emph{runtime verification}
for kernel functions: verifying pre/postcondition contracts at function
boundaries without source modification, for any function the monitored process
executes---including functions the developer did not anticipate.

\subsection{Functions as Black-Box Subsystems}

Every kernel function call is a boundary between a \emph{caller} (who provides
inputs) and a \emph{callee} (a black box from the caller's perspective):
$$\text{Caller} \xrightarrow{[\mathit{arg}_1, \ldots, \mathit{arg}_n]}
  f_{\text{callee}} \xrightarrow{[\mathit{ret}]} \text{Caller}$$

The caller cannot observe what happens inside the callee. It can only choose
arguments (precondition) and observe the return value (postcondition).
\toolname{} captures \emph{both sides} of this boundary, making the black box
transparent without modifying it.

\subsection{Formal Boundary Contracts}

Define a \emph{Boundary Contract} $\mathcal{C}(f) = (\mathcal{P},
\mathcal{Q})$ for function $f$:
\begin{itemize}[nosep]
  \item $\mathcal{P}(\mathit{args})$: precondition predicate over input arguments
  \item $\mathcal{Q}(\mathit{args}, \mathit{ret})$: postcondition predicate
        over return value given inputs
\end{itemize}

\noindent\textbf{Example:} For the Rust binder's
\texttt{Process::set\_max\_threads(max: u32)}, reached from
\texttt{ioctl\_write\_only()} on \texttt{BINDER\_SET\_MAX\_THREADS}:
\begin{align*}
\mathcal{P}&: \mathit{max} \leq \mathit{plausible\_thread\_cap} \\
\mathcal{Q}&: \mathit{inner.max\_threads} = \mathit{max}
\end{align*}

A \emph{contract violation} occurs when the caller passes arguments violating
$\mathcal{P}$ (caller bug) or the callee returns values violating $\mathcal{Q}$
given valid inputs (callee bug). Section~\ref{sec:realworld} reports what
happened when I actually wrote this predicate: the capture was correct and the
predicate was wrong---\texttt{set\_max\_threads()} storing an arbitrary
\texttt{u32} is harmless because \texttt{RLIMIT\_NPROC} is enforced elsewhere.
A contract database is only as good as its predicates, which is why the tool
reports evidence and not verdicts.

\subsection{The Unknown Kernel Path Problem}

When a user-space process issues
\texttt{ioctl(binder\_fd, BINDER\_WRITE\_READ, \&bwr)}, it triggers a chain of
kernel functions---all of them Rust in this tree, and none of them reachable by
\texttt{write()}, since the Rust binder registers no \texttt{write} file
operation at all:

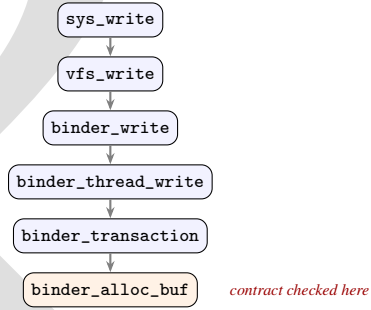
\begin{center}
\begin{tikzpicture}[
  node distance=0.25cm,
  call/.style={rectangle, draw, rounded corners, fill=blue!5, minimum height=0.45cm, font=\ttfamily\scriptsize, inner sep=3pt},
  arrow/.style={-{Stealth[length=1.5mm]}, thick, gray},
]
\node[call] (s1) {sys\_ioctl};
\node[call, below=of s1] (s2) {rust\_binder\_ioctl};
\node[call, below=of s2] (s3) {Process::ioctl};
\node[call, below=of s3] (s4) {Process::ioctl\_write\_read};
\node[call, below=of s4] (s5) {Thread::write\_read};
\node[call, below=of s5] (s6) {Thread::write};
\node[call, below=of s6, fill=orange!10] (s7) {range\_alloc::allocate};
\draw[arrow] (s1) -- (s2);
\draw[arrow] (s2) -- (s3);
\draw[arrow] (s3) -- (s4);
\draw[arrow] (s4) -- (s5);
\draw[arrow] (s5) -- (s6);
\draw[arrow] (s6) -- (s7);
\node[font=\tiny\itshape, right=0.3cm of s7, text=red!60!black] {contract checked here};
\node[font=\tiny\itshape, right=0.3cm of s2, text=black!55] {C ABI shim};
\node[font=\tiny\itshape, right=0.3cm of s3, text=black!55] {mangled Rust symbols below here};
\end{tikzpicture}
\end{center}

Everything below the \texttt{extern "C"} shim carries a mangled Rust symbol, so
naming these boundaries for a kprobe or an ftrace filter means knowing the
mangling, and decoding their arguments means decoding Rust types.
The user cannot predict which functions will execute. Linux
RV~\cite{dealmeida2019rv} requires
pre-defining which events to monitor. \toolname{} captures \emph{every}
function boundary automatically---the offline verifier then checks each
captured $(\mathit{args}, \mathit{ret})$ pair against a contract database.

\subsection{Comparison with Linux RV}

\begin{table}[t]
\centering
\caption{Linux RV vs.\ \toolname{} contract verification.}\label{tab:rv}
\small
\begin{tabular}{@{}lcc@{}}
\toprule
\textbf{Dimension} & \textbf{Linux RV} & \textbf{\toolname{}} \\
\midrule
Scope & Pre-defined events & Any function executed \\
Model & Automata on sequences & Predicates on values \\
Instrumentation & Explicit tracepoints & Zero source modification \\
Catches & Ordering violations & Value violations \\
Non-deterministic state & No (static model) & Yes (runtime values) \\
Unknown functions & Cannot verify & Captures automatically \\
Rust support & No & Yes \\
Specification & Hardcoded in kernel & External database \\
Compile-time sufficiency & N/A & No (value depends on runtime) \\
Overhead & Near zero (event-driven) & 18--21\,ns per record \\
\bottomrule
\end{tabular}
\end{table}

Table~\ref{tab:rv} summarizes the differences. Linux RV verifies
\emph{anticipated} event orderings (e.g., ``sleep must not occur in atomic
context''). \toolname{} verifies \emph{arbitrary} value contracts (e.g.,
``\texttt{max\_threads} must be within a plausible cap'') on functions the
developer never explicitly instrumented.

\subsection{Concrete Example: Detecting the errno Bug}

Contract for \texttt{Error::from\_errno(errno)}:
$$\mathcal{P}: -4095 \leq \mathit{errno} \leq -1$$
$$\mathcal{Q}: \mathit{ret.code} = \mathit{errno}$$

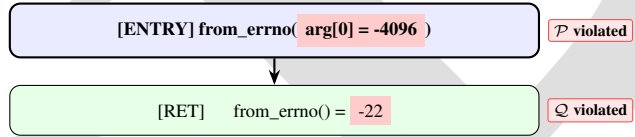
\toolname{} capture:

\begin{center}
\begin{tikzpicture}[
  node distance=0.3cm,
  funcbox/.style={rectangle, draw, thick, rounded corners, fill=blue!8, minimum width=7cm, minimum height=0.5cm, font=\scriptsize\bfseries, align=left, inner sep=4pt},
  retbox/.style={rectangle, draw, rounded corners, fill=green!10, minimum width=7cm, minimum height=0.5cm, font=\scriptsize, align=left, inner sep=4pt},
  violabel/.style={rectangle, draw=red!70, fill=red!10, font=\tiny\bfseries, inner sep=2pt, rounded corners=1pt},
  arrow/.style={-{Stealth[length=2mm]}, thick},
]
\node[funcbox] (entry) {[ENTRY] from\_errno(\colorbox{red!20}{arg[0] = -4096})};
\node[violabel, anchor=west] at ([xshift=3pt]entry.east) {$\mathcal{P}$ violated};
\node[retbox, below=0.35cm of entry] (ret) {[RET]\quad\; from\_errno() = \colorbox{red!20}{-22}};
\node[violabel, anchor=west] at ([xshift=3pt]ret.east) {$\mathcal{Q}$ violated};
\draw[arrow] (entry) -- (ret);
\end{tikzpicture}
\end{center}

Verifier output: precondition violated ($-4096 \notin [-4095, -1]$) and
postcondition violated ($\mathit{ret} = -22 \neq -4096$). Linux RV cannot
catch this: no tracepoint exists inside \texttt{from\_errno}, and it is a Rust
function invisible to C-based tracing infrastructure.

\subsection{Struct Member Contract Verification}\label{sec:struct-contract}

Beyond scalar argument bounds, \toolname{}'s IR-verified pointee snapshots (for
\texttt{byval}, Rust \texttt{\&T}/\texttt{\&mut T}, or \texttt{[static N]}
parameters, decoded into struct members offline via DWARF) and
\texttt{0xA} pointer-access tracking (for plain C pointers) enable
\emph{struct member contracts}: predicates over relationships between fields
that must hold for a function's internal logic to operate correctly.

\paragraph{The problem.} Most kernel functions receive struct pointers or
references whose fields encode complex state (flags, sizes, pointers, reference
counts). The function's internal logic assumes these fields satisfy certain
relationships (e.g., \texttt{buf\_len $\geq$ data\_offset + data\_size}). These
relationships are typically enforced by conditional checks inside the function,
but:
\begin{itemize}[nosep]
  \item Checks may be missing for uncommon field combinations.
  \item Concurrent modification may invalidate a field between the check and
        its use (TOCTOU).
  \item The caller may construct a struct in an inconsistent state that passes
        individual field validation but violates cross-field invariants.
\end{itemize}

Compile-time type systems cannot detect these: the struct is well-typed
regardless of whether its fields are internally consistent. The correct state
depends on runtime conditions (which allocator path was taken, whether a lock
is held, what RCU grace period is active).

\paragraph{Formalization.} Extend the boundary contract to include a
\emph{struct consistency predicate} $\mathcal{S}$ over fields of a struct
parameter:
$$\mathcal{S}(\mathit{fields}): \bigwedge_i \phi_i(f_1, f_2, \ldots, f_n)$$

\noindent\textbf{Example:} For an \texttt{smb2\_create\_req *req} passed with a
static/IR contract or observed via \texttt{0xA} field loads:
\begin{align*}
\mathcal{S}&: \mathit{req.NameLength} \leq \mathit{req.BufferLength} \\
           &\wedge\ \mathit{req.CreateContextsOffset} + \mathit{req.CreateContextsLength}
              \leq \mathit{req.StructureSize}
\end{align*}

\paragraph{Detection mechanism.} Where an argument carries an IR memory or
dereferenceability contract (\texttt{byval}, Rust \texttt{\&msg}, or
\texttt{msg[static 1]}), \toolname{} captures its entry pointee bytes in one
record (Figure~\ref{fig:struct-verify}); for an unannotated C pointer,
\texttt{0xA} access records capture each field loaded during execution. Using
DWARF layout metadata offline (\texttt{sancov -symbolize-dataflow}), the
verifier reconstructs member values and evaluates $\mathcal{S}$. A violation
means the function received inconsistent struct state---either constructed
incorrectly by the caller or modified concurrently.

\begin{figure}[t]
\centering
\begin{tikzpicture}[
  node distance=0.5cm,
  funcbox/.style={rectangle, draw, thick, rounded corners, fill=blue!8, minimum width=6cm, minimum height=0.6cm, font=\scriptsize\bfseries, align=center},
  fieldbox/.style={rectangle, draw, fill=yellow!15, minimum width=5.6cm, font=\ttfamily\scriptsize, align=left, inner sep=4pt},
  instlabel/.style={rectangle, draw=green!60!black, fill=green!10, font=\tiny\bfseries, inner sep=2pt, rounded corners=1pt},
  checkbox/.style={rectangle, draw, thick, fill=orange!10, rounded corners, minimum width=5cm, minimum height=0.6cm, font=\scriptsize, align=center},
  failbox/.style={rectangle, draw, thick, fill=red!15, rounded corners, minimum width=5cm, minimum height=0.6cm, font=\scriptsize\bfseries, align=center},
  globalbox/.style={rectangle, draw, dashed, fill=gray!8, minimum width=2.5cm, font=\scriptsize\itshape, align=center, inner sep=4pt},
  arrow/.style={-{Stealth[length=2mm]}, thick},
]
\node[funcbox] (func) {f(struct msg *m) \textnormal{\scriptsize(\texttt{m[static 1]} / \texttt{\&msg} or \texttt{0xA} loads)}};
\node[instlabel, anchor=south east] at ([xshift=-2pt, yshift=2pt]func.north east) {instrumented};
\node[fieldbox, below=0.8cm of func] (fields) {%
  m->hdr\_len\quad\;\; = 20\\
  m->payload\_len = 4096\\
  m->total\_len\quad = 100\\
  m->flags\qquad\;\;\;\, = 0x3};
\node[instlabel, anchor=north east] at ([xshift=-2pt, yshift=-2pt]fields.south east) {snapshot / 0xA + DWARF};
\node[checkbox, below=1.0cm of fields] (check) {$\mathcal{S}$: hdr\_len + payload\_len $\leq$ total\_len ?};
\node[failbox, below=0.6cm of check] (fail) {20 + 4096 = 4116 $>$ 100 $\to$ VIOLATION};
\draw[arrow] (func) -- node[left, font=\tiny\itshape, text=black!60] {snapshot or 0xA} (fields);
\draw[arrow] (fields) -- node[left, font=\tiny\itshape, text=black!60] {offline verifier} (check);
\draw[arrow] (check) -- (fail);
\node[globalbox, right=0.8cm of check] (global) {global state\\(not captured\\unless passed\\as argument)};
\draw[dashed, gray, -{Stealth[length=1.5mm]}] (func.east) -- ++(1.2,0) |- (global.north);
\end{tikzpicture}
\caption{Struct member contract verification. An IR-verified entry snapshot
(\texttt{m[static 1]} / Rust \texttt{\&msg}) or \texttt{0xA} access stream
records the pointee bytes/values, and the offline verifier cuts them into DWARF
fields to check cross-field predicates.}\label{fig:struct-verify}
\end{figure}

\paragraph{Concrete example.} Consider a kernel network handler:
\begin{lstlisting}[numbers=none]
[ENTRY/ACCESS] process_msg(struct{.hdr_len=20, .payload_len=4096,
                                  .total_len=100, .flags=0x3})
\end{lstlisting}
Contract: $\mathit{hdr\_len} + \mathit{payload\_len} \leq \mathit{total\_len}$.
Captured: $20 + 4096 = 4116 > 100$ --- violation. The function will either
read out-of-bounds or silently produce incorrect output. No crash occurs
(the buffer may be large enough by coincidence), no KASAN report fires,
and edge coverage sees the same path regardless of field values.

\paragraph{Why this matters.} These bugs are among the hardest to find:
\begin{itemize}[nosep]
  \item They produce no immediate crash (data is read, just incorrectly).
  \item Sanitizers do not fire (the access is technically in-bounds of the
        allocation, just semantically wrong).
  \item Static analysis cannot reason about runtime struct state.
  \item They often manifest only under specific timing or workload conditions.
\end{itemize}
Struct member contract verification with \toolname{} detects these violations
at runtime with zero source annotation, by comparing the actual member values
against the cross-field predicates that the function's conditional logic
implicitly assumes. An eBPF program with a hand-written per-function reader can
reach the same values; what \toolname{} removes is the per-function kernel
instrumentation work, since the compiler emits the raw snapshot or \texttt{0xA}
access records automatically and offline DWARF provides the member layout.

\paragraph{Formal model.}
Let a function $f$ receive a struct parameter $s$ with fields
$(f_1, \ldots, f_n)$. Define the \emph{structural verification predicate}:
$$\mathcal{V}(f) = \mathcal{S}(f_1, \ldots, f_n)$$
where $\mathcal{S}$ encodes cross-field invariants that must hold for the
function's logic to operate correctly
(e.g., $\mathit{len} \leq \mathit{cap}$,
$\mathit{offset} + \mathit{size} \leq \mathit{total}$).

\noindent A \emph{violation} $\neg\mathcal{V}(f)$ indicates the
function operated on a struct violating its invariant.

\paragraph{Scope of capture.} Under the v5 IR-only contract, entry snapshots
cover one dereference level on IR-\texttt{dereferenceable} pointers and
\texttt{byval}/\texttt{sret} objects, while \texttt{0xA} records observe direct
constant-offset accesses on plain C pointers (Section~\ref{sec:deref}).
Global kernel state (e.g., slab pressure, lock state) is not captured unless
passed explicitly. The predicate $\mathcal{S}$ therefore checks invariants
within parameters reachable under these two mechanisms.

\paragraph{Classification of violations.}
\begin{center}
\small
\begin{tabular}{@{}lll@{}}
\toprule
\textbf{Violation type} & \textbf{Predicate} & \textbf{Consequence} \\
\midrule
Cross-field overflow & $\mathit{off} + \mathit{len} > \mathit{cap}$ & OOB read/write \\
Length inconsistency & $\mathit{hdr} + \mathit{payload} > \mathit{total}$ & Truncation/overread \\
Flag-field mismatch & $\mathit{flags} \wedge F \neq 0 \wedge \mathit{ptr} = 0$ & NULL deref \\
Size exceeds bound & $\mathit{req\_sz} > \mathit{max\_allowed}$ & Allocation overflow \\
\bottomrule
\end{tabular}
\end{center}

\subsection{Toward Post-Mortem Formal Verification}\label{sec:postmortem}

Classical formal verification proves properties for \emph{all} inputs
statically, but is intractable for 30M~LOC kernels. I observe that
\toolname{}'s structured output enables a complementary approach:
\emph{post-mortem contract verification}---mechanically checking recorded
execution against formal specifications after the fact.

\paragraph{Key Insight.} The record buffer produces a complete sequence of
$\langle \mathit{PC}, \mathit{args}, \mathit{ret} \rangle$ tuples for every
instrumented function the process executed. This is a \emph{finite execution
trace} in the formal methods sense---the input to a runtime verification
monitor.

\paragraph{What This Enables.}
\begin{enumerate}[nosep]
  \item \textbf{Offline model checking:} Replay the buffer against temporal
        logic specifications (e.g., LTL: ``every \texttt{alloc} is eventually
        followed by \texttt{free} with the same pointer'').
  \item \textbf{Contract refinement:} Automatically infer likely contracts from
        many executions (Daikon-style invariant detection on kernel functions),
        then flag violations in new runs.
  \item \textbf{Differential verification:} Compare argument/return value
        distributions across kernel versions to detect semantic regressions
        that pass all tests but violate implicit contracts.
\end{enumerate}

\paragraph{Soundness Boundary.} This is \emph{not} full formal verification:
it only covers paths actually exercised. However, combined with a fuzzer that
maximizes path diversity, it approaches bounded model checking---verifying
contracts over all \emph{reachable} states discovered during exploration.
The structured, machine-readable output (unlike \texttt{printk} logs) makes
this amenable to automated reasoning tools without manual log parsing.

\section{Discussion \& Limitations}\label{sec:discussion}

\subsection{Runtime-Originated Kernel Code Auditing}\label{sec:workflow}

I define \emph{runtime-originated auditing} as the methodology in which the
analysis target is not the source code itself, but the dataflow records captured
during actual kernel execution triggered by a user-space operation. This inverts
the conventional workflow and eliminates the speculative gap inherent in
source-only review (Table~\ref{tab:methodology}, Figure~\ref{fig:workflow}).

\begin{table}[t]
\centering
\caption{Source-speculative vs.\ evidence-based auditing.}\label{tab:methodology}
\small
\begin{tabular}{@{}lll@{}}
\toprule
\textbf{Dimension} & \textbf{Without \toolname{}} & \textbf{With \toolname{}} \\
\midrule
Input & Source code & Captured records \\
Evidence & ``Path appears reachable'' & PC + arg values at runtime \\
Validation & Crash or sanitizer required & Record is the evidence \\
Reproducibility & Analyst-dependent & Deterministic artifact \\
Scope & Anticipated paths only & Every boundary executed \\
\bottomrule
\end{tabular}
\end{table}

\begin{figure*}[t]
\centering
\begin{tikzpicture}[
  node distance=0.35cm,
  box/.style={rectangle, draw, rounded corners, minimum width=2.8cm, minimum height=0.65cm, font=\scriptsize, align=center},
  arrow/.style={-{Stealth[length=2mm]}, thick},
  title/.style={font=\scriptsize\bfseries}
]
\node[title, text=red!70!black] (lt) at (0,0) {Existing: KASAN + KCOV + Fuzzer};
\node[box, fill=red!5, below=0.3cm of lt] (l1) {Fuzzer mutates syscalls};
\node[box, fill=red!5, below=of l1] (l2) {KCOV: edge coverage\\(which paths hit?)};
\node[box, fill=red!5, below=of l2] (l3) {KASAN: memory error\\(crash or no signal)};
\node[box, fill=red!5, below=of l3] (l4) {Coverage saturates\\on stateful subsystems};
\draw[arrow] (l1) -- (l2);
\draw[arrow] (l2) -- (l3);
\draw[arrow] (l3) -- (l4);
\node[font=\tiny\itshape, text=gray, below=0.1cm of l4] {blind to argument values};

\node[title, text=green!50!black] (mt) at (5.5,0) {\toolname{}};
\node[box, fill=green!5, below=0.3cm of mt] (m1) {Same syscall triggers};
\node[box, fill=green!8, below=of m1] (m2) {Capture (PC, args, ret)\\at every IR boundary};
\node[box, fill=green!8, below=of m2] (m3) {IR objects/refs +\\optional \texttt{0xA} accesses};
\node[box, fill=green!5, below=of m3] (m4) {Value-aware feedback\\breaks saturation};
\draw[arrow] (m1) -- (m2);
\draw[arrow] (m2) -- (m3);
\draw[arrow] (m3) -- (m4);
\node[font=\tiny\itshape, text=gray, below=0.1cm of m4] {observes what drove the path};

\node[title, text=blue!70!black] (rt) at (11,0) {Combined Workflow};
\node[box, fill=blue!5, below=0.3cm of rt] (r1) {Fuzzer + edge coverage\\(explore structure)};
\node[box, fill=blue!5, below=of r1] (r2) {\toolname{}: argument values\\(explore state space)};
\node[box, fill=blue!5, below=of r2] (r3) {KASAN/KMSAN: detect\\corruption when triggered};
\node[box, fill=blue!8, below=of r3] (r4) {Root-cause from dataflow:\\which arg $\to$ which field};
\draw[arrow] (r1) -- (r2);
\draw[arrow] (r2) -- (r3);
\draw[arrow] (r3) -- (r4);
\node[font=\tiny\itshape, text=gray, below=0.1cm of r4] {full observability stack};

\draw[dashed, gray] (2.8,0.2) -- (2.8,-4.5);
\draw[dashed, gray] (8.2,0.2) -- (8.2,-4.5);
\end{tikzpicture}
\caption{Left: existing kernel fuzzing stack (edge-blind to values).
Middle: \toolname{} adds boundary data-flow extraction.
Right: combined workflow provides full observability from mutation guidance
through root-cause analysis.}\label{fig:workflow}
\end{figure*}

\begin{figure}[t]
\centering
\begin{tikzpicture}[
  node distance=0.6cm,
  box/.style={rectangle, draw, thick, rounded corners, minimum width=3.2cm, minimum height=0.7cm, font=\scriptsize, align=center},
  arrow/.style={-{Stealth[length=2mm]}, thick},
]
\node[box, fill=yellow!15] (cap) {\toolname{} capture\\(PC, args, ret, struct fields)};

\node[box, fill=green!12, below left=1cm and 0.3cm of cap] (audit) {Security Audit\\contract violations};
\node[box, fill=red!10, below right=1cm and 0.3cm of cap] (fuzz) {Guided Fuzzing\\value-aware mutation};

\node[box, fill=blue!10, below=2.5cm of cap] (result) {Confirmed finding\\reproducible evidence};

\draw[arrow] (cap.south) -- ++(0,-0.25) -| (audit.north);
\draw[arrow] (cap.south) -- ++(0,-0.25) -| (fuzz.north);
\draw[arrow] (audit.south) -- ++(0,-0.25) -| (result.north);
\draw[arrow] (fuzz.south) -- ++(0,-0.25) -| (result.north);

\node[font=\tiny\itshape, text=black!60] at (-2.2,-1.0) {offline};
\node[font=\tiny\itshape, text=black!60] at (2.2,-1.0) {online};
\end{tikzpicture}
\caption{Dual utility: the same captured records serve both manual security
auditing (offline) and automated fuzzing guidance (online), converging on
confirmed findings with reproducible evidence.}\label{fig:dual}
\end{figure}

\paragraph{Key property.} The dataflow records are a \emph{deterministic
artifact}: given the same kernel build, the same user action produces the same
records. Any party can independently reproduce and verify the analysis,
satisfying the reproducibility requirement for empirical security research.

\paragraph{Dual utility.} As Figure~\ref{fig:dual} shows, the same records
serve both (1)~security auditing
(identifying argument values that violate API contracts) and (2)~guided fuzzing
(constraining mutation to value ranges observed at deep kernel code paths,
eliminating blind 64-bit enumeration).

\paragraph{Implication for AI-assisted auditing.} Without runtime evidence,
AI-based code review produces hypotheses that may be incorrect (as demonstrated
by my initial binder claims refuted by maintainers). With \toolname{}, the AI
analyzes observed records rather than speculating about behavior---findings are
grounded in physical execution regardless of reasoning quality.

\paragraph{IR-Only Invariance and Pointer Visibility.} Instrumentation emits
identical callbacks, operands, and machine code with or without \texttt{-g};
DWARF (\texttt{CONFIG\_DEBUG\_INFO}) is used solely for optional offline
symbolization (\texttt{sancov -symbolize-dataflow} / \texttt{addr2line}).
Under this rule, pointee snapshots at function entry are taken only where IR
attributes guarantee readability (\texttt{byval}/\texttt{sret}/\texttt{noundef
dereferenceable}), which leaves plain C \texttt{T *} parameters as addresses
until \texttt{CONFIG\_KCOV\_DATAFLOW\_POINTEE\_ACCESS} (\texttt{0xA}) records
loads and stores through them at runtime.

\paragraph{Rust Pipeline.} Dataflow tracking of Rust kernel modules requires
building \texttt{rustc} against the custom LLVM~24 that contains the
trace-args/ret passes. The post-compilation pipeline (Path~A) serves as a
proof-of-concept demonstrating language-agnosticism but cannot instrument
modules built through Kbuild. I propose adding native
\texttt{-Zsanitizer-coverage=trace-args} support to upstream rustc as future
work to eliminate the custom build requirement.

\paragraph{Buffer Overflow and Remote Pool Bounds.} If instrumented code
generates more events than the buffer capacity, events are silently dropped
without sleeping, allocating, or locking. In v5, \texttt{kcov\_df\_reserve()}
and \texttt{kcov\_df\_merge()} increment per-CPU counters exported via read-only
debugfs (\texttt{kcov\_dataflow\_dropped\_records} and
\texttt{kcov\_dataflow\_dropped\_sections}), while inert-context checks remain
uncounted so idle \texttt{INSTRUMENT\_ALL} paths pay no RMW penalty. Bounded
scratch sizing (\texttt{KCOV\_DF\_REMOTE\_ENABLE\_ARG}), an atomic epoch fence
(\texttt{KCOV\_DF\_REMOTE\_RESET}), a page-unit shrinker, and a 64\,MiB idle
pool cap in \texttt{kernel/kcov.c} prevent remote kworker storms from exhausting
\texttt{vmalloc} space.

\paragraph{Interrupt, NMI, and nofault Context.} Instrumentation callbacks bail
out when \texttt{!in\_task() || pagefault\_disabled()}: a call originating from
a softirq, hardirq, or NMI handler is silently discarded, and so is one issued
from inside a \texttt{copy\_from\_kernel\_nofault()} region.
The first half is stricter than the per-context filtering used by mainline KCOV
(which allows softirq tracing via \texttt{kcov\_remote\_start}); I chose
the stricter policy because the dataflow callback performs pointer
dereferences (\texttt{copy\_from\_kernel\_nofault}) whose latency is
unacceptable in hard-interrupt paths, and because per-task buffer
interleaving from nested contexts would corrupt record boundaries.
The second half is a containment measure rather than a policy: an instrumented
leaf reached from inside the collector's own nofault read---or from a KASAN
stack walk---otherwise issues a callback per load and soft-locks the CPU, and
the per-task recursion guard only covers re-entry nested inside our own
callback, not a fresh entry from the unwinder. The trade-off is that records
are also dropped inside unrelated \texttt{pagefault\_disable()} regions
(\texttt{kmap\_atomic()} on HIGHMEM, futex and perf callchain probes), which
are short and rare on the targeted workloads.
Kernel threads are still reachable: \texttt{kcov\_df\_remote\_start()}
annotations let a kworker record into a published remote handle, as mainline
\kcov{} does, but only in sleepable task context---it takes a mutex---so it
closes the kworker gap and not the softirq one.
Bugs triggered exclusively from interrupt context are therefore invisible
to \toolname{}; however, such bugs are also unreachable by KCOV-guided
fuzzers, which exercise kernel code through syscalls in process context.

\paragraph{Fuzzer Integration.} Beyond the \texttt{compiler-rt} weak stubs and
companion libFuzzer value-profile hook (Section~\ref{sec:libfuzzer}),
\toolname{} now powers full-system kernel fuzzing on \texttt{ksmbd} via
per-connection remote handle routing and value-aware coverage folding (alongside
\texttt{0xC} comparison operand solving). Generalizing the same executor
integration upstream into \texttt{syzkaller} remains natural future work.

\paragraph{Upstream Path.} The LLVM pass has been structured as a clean
four-patch upstream stack---the IR-only \sancov{} pass, Clang driver flags,
weak \texttt{compiler-rt} defaults, and \texttt{clang/docs/SanitizerCoverage.md}
specification---associated with the pull
request\footnote{\url{https://github.com/llvm/llvm-project/pull/201410}} and LLVM
Discourse
RFC\footnote{\url{https://discourse.llvm.org/t/rfc-sanitizercoverage-add-fsanitize-coverage-trace-args-trace-ret/91026}},
with \texttt{\_\_sancov\_dataflow} offline symbolization and
\texttt{-sanitizer-coverage-trace-pointee-access} (\texttt{0xA}) developed on
companion follow-on branches. On the Linux side, the RFC patch
series\footnote{\url{https://lore.kernel.org/all/20260611-b4-kcov-dataflow-v2-v2-0-0a261da3987c@est.tech/}}
has been updated to match the IR-only v5 ABI, adding \texttt{0xA} pointer-access
records, \texttt{KCOV\_DF\_REMOTE\_ENABLE\_ARG}, \texttt{KCOV\_DF\_REMOTE\_RESET},
per-CPU drop counters, the page-unit scratch shrinker, and the 64\,MiB idle
remote-area pool cap in \texttt{kernel/kcov.c}.

\section{Related Work}\label{sec:related}

\paragraph{Coverage-Guided Kernel Fuzzing.}
syzkaller~\cite{vyukov2015syzkaller} and kAFL~\cite{schumilo2017kafl} use
edge/block coverage as feedback. DIFUZE~\cite{corina2017difuze} adds
interface-aware generation but no runtime data-flow.
MORPHUZZ~\cite{bulekov2022morphuzz} fuzzes virtual devices without data-flow
context. HEALER~\cite{sun2021healer} and
Rtkaller~\cite{ma2022rtkaller} learn relation models between syscalls to
construct deeper sequences, yet both remain blind to argument values and
plateau once the structural path space is exhausted.
SyzVegas~\cite{wang2021syzvegas} applies multi-armed bandit algorithms to
syzkaller's mutation strategy but still uses edge coverage as the sole reward
signal. ACTOR~\cite{fleischer2023actor} targets concurrency bugs via
schedule-aware fuzzing; its feedback is scheduling order, not data-flow.
Moonshine~\cite{pailoor2018moonshine} distills seeds from system call traces
to improve initial coverage but cannot guide mutation once coverage saturates.
Google's AIxCC kernel fuzzing approach~\cite{google2024aixcc} combines AFL
with KCOV and KASAN in a QEMU harness, reporting that ``the vulnerability
cannot be easily found even after hours of fuzzing, unless provided with seed
inputs close to the solution''---confirming that edge coverage alone saturates
on complex kernel targets.
All of the above rely solely on control-flow coverage; none capture the
argument values that drive state-dependent behavior.

\paragraph{Data-Flow Guided Fuzzing.}
Payer~\cite{payer2019dataflow} argues for data-flow-guided fuzzing but provides
no kernel implementation. Angora~\cite{chen2018angora} uses byte-level taint for
user-space programs at 10$\times$ overhead.
GREYONE~\cite{gan2020greyone} combines taint with coverage but targets
user-space only. Both pay an order-of-magnitude cost for byte-level precision.
\toolname{} makes the opposite trade: it observes values only where they cross
a function boundary, which is far less than full taint but costs a bounded
number of callbacks per call rather than a shadow operation per byte.

\paragraph{Kernel Instrumentation.}
KCOV~\cite{vyukov2016kcov} provides edge coverage---my foundation---and its
\texttt{trace-cmp} mode provides comparison operands, which \toolname{} reuses
rather than duplicates: when both are configured the operand pairs land in the
same buffer as the boundary records. \texttt{ftrace}, \texttt{kprobes}, and
eBPF patch global kernel text, so a probe fires for every task on every CPU and
per-task selection is a filter rather than a property of the instrumentation;
ftrace's argument printing is additionally capped at six parameters and depends
on BTF shapes that idiomatic Rust types do not produce
(Section~\ref{sec:ftrace}).

\paragraph{Post-Mortem Analysis.}
\texttt{drgn}~\cite{drgn} enables programmable vmcore analysis but requires a
crash and fails on optimized Rust code. \texttt{kdump}/\texttt{crash} provide
full memory dumps but require manual analysis and miss silent bugs.

\medskip\noindent
\toolname{} combines compile-time IR-only data-flow instrumentation
with a production-grade kernel transport, supporting both C and Rust, with
v4 prototype measurements establishing 18--21\,ns per recorded value and +66\%
idle syscall overhead under whole-kernel instrumentation
(Section~\ref{sec:eval}), and v5 hardening the ABI, remote collection, and
pointee visibility. Weak \texttt{compiler-rt} hooks and the companion
libFuzzer runtime confirm that the three-operand ABI is a general
\sancov{} mechanism rather than kernel-only.

\section{Conclusion}\label{sec:conclusion}

I presented \toolname{}, bridging the semantic gap between coverage-guided
fuzzing and data-flow analysis for OS kernels. My key insight is that function
boundaries are natural observation points where data-flow context is both cheap
to capture and highly informative for both automated exploration and human
triage.

The system requires no source annotation, operates safely in sleepable task
context with explicit rejection of interrupt, NMI, and nofault callbacks,
reaches kernel threads through a remote-handle path, and---uniquely---decodes
Rust kernel module arguments via both a post-compilation
pipeline and native instrumentation through a custom-built rustc. That last
point is narrower than it may appear: \texttt{ftrace} can place a probe in a
Rust object, but cannot decode idiomatic Rust argument types, and
\texttt{drgn} cannot read them at all once \texttt{-O2} has elided the DWARF
variable locations.

I demonstrated baseline correctness on the in-tree selftest suite covering
in-memory object/struct expansion in C and Rust, arguments past the ABI's
register budget, indirect \texttt{sret} returns, a crashless postcondition
violation, and bounded remote capture from kworker context---augmented in v5 by
strict \texttt{-g}-invariant IR instrumentation, \texttt{KCOV\_DF\_TYPE\_ACCESS}
pointer tracking, and bounded/epoch-fenced remote pools. Weak
\texttt{compiler-rt} hooks and libFuzzer integration confirm that the ABI is not
kernel-specific. Source code is available at \anonrepo{}.

\bibliographystyle{unsrt}
\bibliography{../shared/references}

\appendix
\section{LLVM Pass: IR-Only \texttt{InjectTraceForArgs} Pseudocode}\label{app:pass}

\begin{algorithmic}[1]
\Function{InjectTraceForArgs}{$F$}
  \State $\mathit{IP} \gets$ first legal insertion point in $F$'s entry block (after static allocas)
  \State Attach $F$'s opening \texttt{!dbg} location (if present; location only)
  \For{each IR argument $\mathit{Arg} \in F.\mathrm{args}()$}
    \If{$\mathit{Arg}.\mathrm{hasStructRetAttr}()$ \textbf{or} $\mathit{Arg}.\mathrm{hasSwiftErrorAttr}()$}
      \State \textbf{continue} \Comment{\texttt{sret}/\texttt{swifterror} skipped; IR index $\mathit{idx}$ preserved}
    \EndIf
    \State $\mathit{Unused} \gets \mathit{Arg}.\mathrm{use\_empty}()$
    \State $\mathit{idx} \gets \mathit{Arg}.\mathrm{getArgNo}()$
    \If{$\mathit{MemTy} \gets \mathit{Arg}.\mathrm{getPointeeInMemoryValueType}()$}
      \Comment{\texttt{byval}/\texttt{byref}/\texttt{inalloca}/\texttt{preallocated}}
      \State \Call{TraceObject}{$\mathit{idx}$, $\mathit{Unused}$ ? null : $\mathit{Arg}$, $\mathit{Unused}$ ? null : $\mathit{MemTy}$}
    \ElsIf{$\neg\mathit{Unused}$ \textbf{and} $\mathit{Arg}$ is ptr \textbf{and} $\mathit{Arg}.\mathrm{hasNoUndef}()$ \textbf{and} $B \gets \Call{GetPointeeBytes}{\mathit{Arg}} > 0$}
      \State \Call{TracePointee}{$\mathit{idx}$, $\mathit{Arg}$, $B$}
        \Comment{capped by \texttt{-sanitizer-coverage-trace-pointee-bytes}}
    \Else
      \For{each leaf element $e$ of $\mathit{Arg}$ (unwrapping first-class aggregates)}
        \State $(v, \mathit{sz}) \gets \Call{GetReportedValue}{e}$ \Comment{freezes poison; $\mathit{sz}=0$ if unreportable}
        \State \Call{Emit}{$\mathit{idx}$, $\mathit{sz}$, $v$, $\varnothing$}
      \EndFor
    \EndIf
  \EndFor
\EndFunction
\Statex
\Function{Emit}{$\mathit{idx}$, $\mathit{sz}$, $\mathit{val}$, $\mathit{Off}$}
  \State $\mathit{tag} \gets ((\mathrm{u64})\mathit{idx} \ll 32) \mid (\mathrm{u64})\mathit{sz}$
  \State emit \texttt{notail nomerge} \texttt{trace\_args($\mathit{tag}$, $\mathit{val}$, $\mathit{Off}$)}
\EndFunction
\end{algorithmic}

\noindent Two invariants keep instrumentation strictly IR-only and crash-free:
DWARF metadata is never consulted when deciding which calls or tables to emit
(debug info only attaches a \texttt{!dbg} location when present), and unused
in-memory arguments pass \texttt{null} so dead-argument poison pointers are
never dereferenced. Every non-null \texttt{offsets} table begins with
\texttt{offsets[0] = N}, followed by $N$ \emph{(byte offset, byte size)} pairs
($\leq 8$ bytes each), memoized per IR type in \texttt{FieldOffsetsCache}.

\section{Record Buffer Format}\label{app:ringbuf}

Each record occupies $3 + \mathit{nvals}$ quadwords (64-bit each), where
$\mathit{nvals}$ is $N$ from \texttt{offsets[0]}, 1 for scalar values and
\texttt{ACCESS} records, and 2 for comparisons:

\begin{center}
\small
\begin{tabular}{|c|l|p{10.0cm}|}
\hline
\textbf{Offset} & \textbf{Field} & \textbf{Description} \\
\hline
0 & \texttt{header} & \texttt{arg\_idx} [63:56] | \texttt{size} [55:48] | \texttt{nvals} [47:32] | \texttt{type} [31:28] | \texttt{MEM} [24] | \texttt{seq} [23:0] \\
1 & \texttt{pc} & Call site that produced the record: callback return address $-1$, KASLR offset removed \\
2 & \texttt{meta} & ENTRY/RET: read address when \texttt{KCOV\_DF\_HDR\_MEM} (bit~24) is set, else 0. CMP: comparison type bits. ACCESS: kind (\texttt{STORE}/\texttt{VOLATILE}/\texttt{ATOMIC}/ordering) \\
3..3+\textit{nvals} & \texttt{val[\,]} & Register value, 8-byte snapshot/object chunks, loaded/stored ACCESS value, or two CMP operands; \texttt{KCOV\_DF\_MAGIC\_BAD} (\texttt{0xBADADD85}) on unreadable/zero-size entries \\
\hline
\end{tabular}
\end{center}

\noindent Record types are \texttt{0xE} (entry), \texttt{0xF} (return),
\texttt{0xC} (comparison), and \texttt{0xA} (pointer access). \texttt{size} and
\texttt{arg\_idx} are clamped to 255, \texttt{nvals} to $2^{16}-1$.
\texttt{area[0]} holds the committed word count, so walking
\texttt{pos += 3 + nvals} while \texttt{pos < 1 + area[0]} stays inside the
mapping and skips unfamiliar record types safely. Constant \texttt{offsets}
tables stay in kernel \texttt{.rodata} and never reach user space.

\section{Rust Post-Compilation Pipeline}\label{app:rust}

\begin{lstlisting}[language=bash,caption={Rust module instrumentation pipeline}]
# 1. Emit LLVM IR text
rustc --edition 2021 --emit=llvm-ir -g \
  -o module.ll module.rs --crate-type lib

# 2. Instrument with our opt (custom LLVM 24)
opt -passes=sancov-module \
  -sanitizer-coverage-level=1 \
  -sanitizer-coverage-trace-args \
  -sanitizer-coverage-trace-ret \
  -S module.ll -o module_inst.ll

# 3. Compile to object
llc -filetype=obj -relocation-model=static \
  -code-model=kernel module_inst.ll -o module.o

# 4. Link as normal kernel module
\end{lstlisting}

\section{Ioctl Interface Specification}\label{app:ioctl}

\begin{lstlisting}[language=C,caption={Ioctl definitions}]
struct kcov_df_remote_arg {
    __aligned_u64 handle;     /* kcov_remote_handle() encoding */
    __aligned_u64 area_size;  /* per-section scratch words in [8, 1<<20], or 0 */
};

/* Buffer size is passed by value, in u64 words. */
#define KCOV_DF_INIT_TRACK        _IOR('d', 1, unsigned long)
#define KCOV_DF_ENABLE            _IO('d', 100)
#define KCOV_DF_DISABLE           _IO('d', 101)
/* Takes a pointer to a kcov_remote_handle()-encoded __u64. */
#define KCOV_DF_REMOTE_ENABLE     _IOW('d', 102, __u64)
#define KCOV_DF_REMOTE_DISABLE    _IO('d', 103)
/* Bounded remote scratch sizing and atomic epoch buffer reset. */
#define KCOV_DF_REMOTE_ENABLE_ARG _IOW('d', 104, struct kcov_df_remote_arg)
#define KCOV_DF_REMOTE_RESET      _IO('d', 105)
\end{lstlisting}

\end{document}